\documentclass[fleqn,usenatbib]{mnras}

\usepackage{newtxtext,newtxmath}
\usepackage[T1]{fontenc}

\DeclareRobustCommand{\VAN}[3]{#2}
\let\VANthebibliography\thebibliography
\def\thebibliography{\DeclareRobustCommand{\VAN}[3]{##3}\VANthebibliography}

\usepackage{graphicx}	% Including figure files
\usepackage{amsmath}	% Advanced maths commands
\usepackage[dvipsnames]{xcolor} % Access to extra colors
\usepackage[free-standing-units]{siunitx}
\usepackage{nicefrac}

\DeclareSIUnit{\yrs}{yrs}  % yrs 
\DeclareSIUnit{\yr}{yr}   % yr

\colorlet{darkpink}{pink!50!black}
\colorlet{darkgreen}{green!50!black}

\definecolor{maroon}{rgb}{0.69, 0.19, 0.38}
\newcommand{\refn}[1]{\textcolor{maroon}{\bf (REF NEEDED)}}

\title[Timescales in overstable planetary systems]{Migration and eccentricity damping timescales in overstable resonant planetary systems: hydrodynamic and N-body simulations}

\author[Z. Afkanpour et al.]{
Zahra Afkanpour,$^{1}$\thanks{E-mail: zahra.afkanpour@alumni.um.ac.ir}
Sareh Ataiee,$^{1,2}$\thanks{E-mail: sarehataiee@um.ac.ir}
Anna B. T. Penzlin,$^{3}$
Alexandros Ziampras,$^{3, 4}$
Hadiseh Nabavinezhad,$^{1}$
\newauthor
Christoph Sch\"afer,$^{5}$
Hilke Schlichting,$^{6}$
and Rafael Sfair$^{5, 7, 8}$
\\
$^{1}$Department of Physics, Ferdowsi University of Mashhad, Mashhad, Iran\\
$^{2}$Research Center for Modelling and Computational Sciences, Ferdowsi University of Mashhad, Mashhad, Iran\\
$^{3}$Ludwig-Maximilians-Universit{\"a}t M{\"u}nchen, Universit{\"a}ts-Sternwarte, Scheinerstr.~1, 81679 M{\"u}nchen, Germany\\
$^{4}$Max Planck Institute for Astronomy, Königstuhl 17, 69117 Heidelberg, Germany\\
$^{5}$Institut f\"ur Astronomie und Astrophysik, Universität T\"ubingen, Auf der Morgenstelle 10, 72076 T\"ubingen, Germany\\
$^{6}$Department of Earth, Planetary, and Space Sciences, The University of California, Los Angeles, 595 Charles E. Young Drive East, Los Angeles, CA 90095, USA\\
$^{7}$Grupo de Dinâmica Orbital e Planetologia, São Paulo State University, UNESP, Guaratinguetá, CEP 12516-410, São Paulo, Brazil\\
$^{8}$LIRA, Observatoire de Paris, Université PSL, Sorbonne Université, Université Paris Cité, CY Cergy Paris Université, CNRS, 92190 Meudon, France
}
\date{Accepted XXX. Received YYY; in original form ZZZ}

\pubyear{2025}

\begin{document}
\label{firstpage}
\pagerange{\pageref{firstpage}--\pageref{lastpage}}
\maketitle

%================================================ Abstract=============================================
\begin{abstract}

Tidal interactions between planets and natal discs modify semi-major axes and eccentricities. These effects are commonly represented by migration and eccentricity damping timescales in analytical and N-body models, although they are usually assumed constant. This approximation is uncertain in resonant multi-planet systems, where coupled planet--disc evolution can produce non--linear torque feedback.
We investigate migration and eccentricity damping in two-planet systems undergoing resonant and overstable evolution near the 2:1 mean-motion resonance. Using hydrodynamic simulations with varying disc surface densities, viscosities, and aspect ratios, we measure both timescales from orbital energy and angular momentum exchange between the planets and the disc.
We find that both timescales are time-dependent and vary throughout the evolution. In resonant systems, sign reversals occur owing to eccentricity oscillations and disc torque variations. 
%Moreover, migration timescales normalised by their expected dependence on planetary mass, disc surface density, and disc aspect ratio deviate between models, indicating additional effects from disc structure and gap formation. 
Moreover, migration timescales normalised by their expected dependence on planetary mass, disc surface density, and aspect ratio deviate between models, indicating additional effects from the evolving planet-disc interaction.
Likewise, the common relation between migration and eccentricity damping timescales is not recovered.
We compare the hydrodynamic evolution with several N-body prescriptions. Models with constant migration and damping timescales fail to reproduce the hydrodynamic evolution, whereas torque-based prescriptions provide better agreement. In particular, prescriptions including co-rotation torque saturation and planetary eccentricity reproduce the hydrodynamic evolution in the type-I migration regime, whereas additional corrections are required when planets partially open gaps.
These results show that time-dependent torque-based prescriptions provide a consistent physical link between hydrodynamic simulations and simplified N-body models of resonant planetary systems.
\end{abstract}

%======================================================================================================
%================================================keywords===============================================
%======================================================================================================

\begin{keywords}
hydrodynamics -- methods: numerical -- software: simulations -- planets and satellites: dynamical evolution and stability -- planet-disc interactions -- protoplanetary discs % \byzahra{I checked with List of astronomical key words (Updated February 2025), one to six key words from this list should write as keywords. CHECKED}
\end{keywords}

%======================================================================================================
%==============================================Introduction============================================
%======================================================================================================

\section{Introduction}

Planet formation and migration occur within protoplanetary discs, where gravitational interactions between planets and the surrounding gas continuously exchange angular momentum and energy. These interactions drive orbital migration and damp planetary eccentricities, thereby shaping the final architecture of planetary systems. 

The theoretical foundations of planet--disc interactions were established by \cite{1980ApJ...GT} and \cite{ ward1997protoplanet}. Since then, the theory of planetary migration and disc-driven orbital evolution has been developed through analytical, numerical, and hydrodynamic studies \citep[see][for reviews]{kley2012planet, 2014prpl.conf..667B}. In particular, \cite{2002ApJ...565.1257T} and \cite{2004ApJ...TW} derived prescriptions for type-I migration and eccentricity damping.

Because long-term hydrodynamic simulations remain computationally expensive, many investigations of planetary system formation rely on N-body calculations in which the effects of the disc are represented through prescribed migration and eccentricity damping timescales, $\tau_a$ and $\tau_e$. Such approaches have been extensively used to study resonant capture, resonant chains, and the long-term dynamical evolution of planetary systems \citep[e.g.][]{lee2002dynamics, papaloizou2005migration, 2014AJ...GS, 2015AJ....149..167B, 2023ApJ...946L..11B}. In most applications, these timescales are assumed to be constant or are parameterised using simple analytical relations. Although this approximation has proven useful, its validity has rarely been investigated against hydrodynamic simulations of resonant multi-planet systems.

The reliability of these prescriptions is particularly important for planets undergoing convergent migration toward mean-motion resonances.  Studies have shown that resonance capture results from a delicate balance between resonant excitation and disc-driven damping \citep{2004A&A...414..735K, papaloizou2005migration, 2015A&A...579A.128D}. Consequently, even modest changes in the migration or eccentricity damping rates can modify the resonant equilibrium, alter the amplitude of libration, and affect whether a resonant configuration remains stable over long timescales \citep{2014AJ...GS, deck2015migration, xu2018migration}.

In resonant systems, gravitational interactions between the planets continuously excite planetary eccentricities, whereas disc--planet interactions usually damp them, unless there is a depletion of gas around the planets' orbits such as a clean gap or surface density drop~\citep{1980ApJ...GT, 2001A&A...366..263P, 2015ApJ...812...94D}. The competition between these prcesses establishes an equilibrium eccentricity. Under certain conditions, however, perturbations about this equilibrium do not decay but instead grow with time, eventually leading to resonance escape. This behaviour is commonly referred to as overstability, was investigated analytically by \cite{2014AJ...GS} and subsequently extended by \cite{deck2015migration}, \cite{2015A&A...579A.128D}, and \cite{xu2018migration}. 

The migration and eccentricity damping rates themselves have also been the subject of extensive investigation. Hydrodynamic studies demonstrated that the torques acting on eccentric planets differ substantially from those predicted for circular orbits and depend on the disc response, spiral-wave structure, and eccentricity evolution \citep{2000MNRAS...PL, 2007A&A...CDKN, 2008A&A...CN}. In addition, torque formulae accounting for Lindblad and corotation torques have revealed a complex dependence on disc structure and thermodynamics \citep{2010MNRAS.401.1950P, 2011MNRAS.410..293P}. These results suggest that migration and eccentricity damping may not always be accurately represented by fixed timescales.

In a previous study, \cite{2024A&A...Zahra} investigated the onset of overstability in resonant two--planet systems using hydrodynamic simulations and identified the disc conditions under which temporary resonant capture occurs. That work focused on the dynamical outcomes of the systems, including resonance capture, overstability, and resonance breaking. However, the migration and eccentricity damping timescales underlying these evolutionary pathways were not examined. 
As a result, it remains unclear whether the damping prescriptions commonly adopted in N-body studies accurately represent the disc-driven evolution of resonant systems close to the onset of overstability.
More generally, despite the widespread use of $\tau_a$ and $\tau_e$ in analytical and N-body studies, systematic measurements of these quantities in resonant multi--planet hydrodynamic simulations remain limited.

This work aims to investigate migration and eccentricity damping timescales in resonant planetary systems that exhibit overstable behaviour and to evaluate the assumptions commonly adopted in N-body models. 
Using hydrodynamic simulations of two planets near the 2:1 mean-motion resonance, we measure these timescales directly from the exchange of angular momentum, orbital energy, and disc torques following \cite{2021A&A...AK}. 
We then address two questions: (i) whether commonly used analytical scalings reproduce the measured hydrodynamic behaviour, and (ii) which N-body prescriptions provide the most faithful representation of the orbital evolution observed in the simulations.
Establishing such a connection is essential if simplified N-body models are to be used reliably in studies of resonant capture, resonant stability, and overstability.

This paper is organised as follows. Section 2 describes the numerical setup and the method used to calculate migration and eccentricity damping timescales. Section 3 presents the hydrodynamic results and parameter study. Section 4 compares the hydrodynamic simulations with different N-body prescriptions. Section 5 discusses the physical interpretation and implications of the measured timescales. Finally, Section 6 summarises our conclusions.

%=======================================================================================================
%===============================================Methods=================================================
%=======================================================================================================

\section{Methods} \label{sec:methods}

In this section, we describe the setup used for hydrodynamic %and N-body 
simulations, as well as the methodology for calculating the disc's effect on migration and eccentricity damping timescales.
%=======================================================================================================%
\subsection{Hydrodynamic setup} \label{subsec:Hydro}

We adopted the hydrodynamic setup and physical model from our previous study \cite{2024A&A...Zahra} (hereafter Af24; see our Section 2 for full details). 

Briefly, our system consisted of two super--Earth planets ($5\, M_{\earth} $ and $10\, M_{\earth} $) orbiting a solar--mass host star. To reduce computational costs, we initially placed the planets near the 2:1 mean--motion resonance at 1.23 $AU$ and 2.0 $AU$, respectively. Depending on the disc's properties, migration could lead to resonance capture.

For the fiducial model, we used a protoplanetary disc with a surface density, $\Sigma(r)=\Sigma_0 (r/r_0)^{-1}$ where $\Sigma_0 = 1.13 \times 10^{-4} M_{\star}/r_0^2$ and $r_0$ is taken unity. A viscosity parameter, $\alpha_{\nu}=10^{-5}$ and an aspect ratio, $h=h_0 (r/r_0)^{0.25}$ with $h_0=0.05$ are used. 

%========================================================================================================%
\subsection{Calculating the migration and eccentricity damping timescales} \label{subsec:how to calculate time} % used for referring to this section from elsewhere

To quantify the disc's effect on planetary migration and eccentricity damping, we used the equations based on the power and torque analysis of \cite{2021A&A...AK}. 
The migration timescale
\footnote{Here, the quantity $\tau_a$ (formally the timescale for the evolution of the semi-major axis) is referred to as the 'migration timescale' for simplicity. Note, however, that it is half of the standard migration timescale $\tau_{\mathrm{mig}}$, i.e. $\tau_{a} = \tfrac{1}{2} \tau_{\mathrm{mig}}$.}
and the eccentricity damping timescale are defined as,
% ====================================== eq: timescales ============================================
\begin{equation} \label{eq:timescales}
\begin{aligned}
\frac{1}{\tau_{a}} &= \frac{\dot{a}}{a} = \frac{P}{|E|}, \\
%T (r) &= T_0 \left(\frac{r}{r_0}\right)^{-\beta}, \\
\frac{1}{\tau_{e}} &= \frac{\dot{e}}{e} = (\frac{1}{2 \tau_{a}} - \frac{\Gamma}{L}) \frac{1-e^2}{e^2},
\end{aligned} 
\end{equation}
%===================================================================================================
where $P$ and $\Gamma$ denote the power and torque exerted by the disc on the planet, respectively, which computed by \texttt{Fargo3D} during the simulations.  
The orbital energy ($E$) and angular momentum ($L$) of the planet are given by,
% =============================== eq: Energy & Angular momentum ======================================
\begin{equation} \label{energy and angular momentum}
\begin{aligned}
E &= - \frac{G M_{\star} m_{p}}{2 a}, \\
L &= m_{p}\sqrt{G M_{\star} a (1-e^2)},
\end{aligned} 
\end{equation}
% ===================================================================================================
where $a$ and $e$ are the semi-major axis and eccentricity of the planet, and $M_{\star}$ and $m_{p}$ are the stellar and planetary masses. In the adopted unit system, we set $G M_{\star} = 1$, and all quantities were normalised by the planet mass.

The direct extraction of timescales from the hydrodynamical simulations presented significant challenges due to strong oscillations in the measured values. 
Conventional smoothing techniques, including Fast Fourier transforms, sliding averages, and the Savitzky–-Golay filter, which have been applied in single--planet studies \citep[e.g.][]{2014MNRAS.437...96F}, proved ineffective for our two--planet system. 
%As shown in Fig.~\ref{fig:tau_ae_column}, 
These methods risk producing physically misleading results due to the non-uniform patterns in the timescale data. For this reason, we did not apply any data smoothing procedures.

% ======================================figure: tau_ae_column==========================================
\begin{figure} 
	
	\includegraphics[width=\columnwidth]{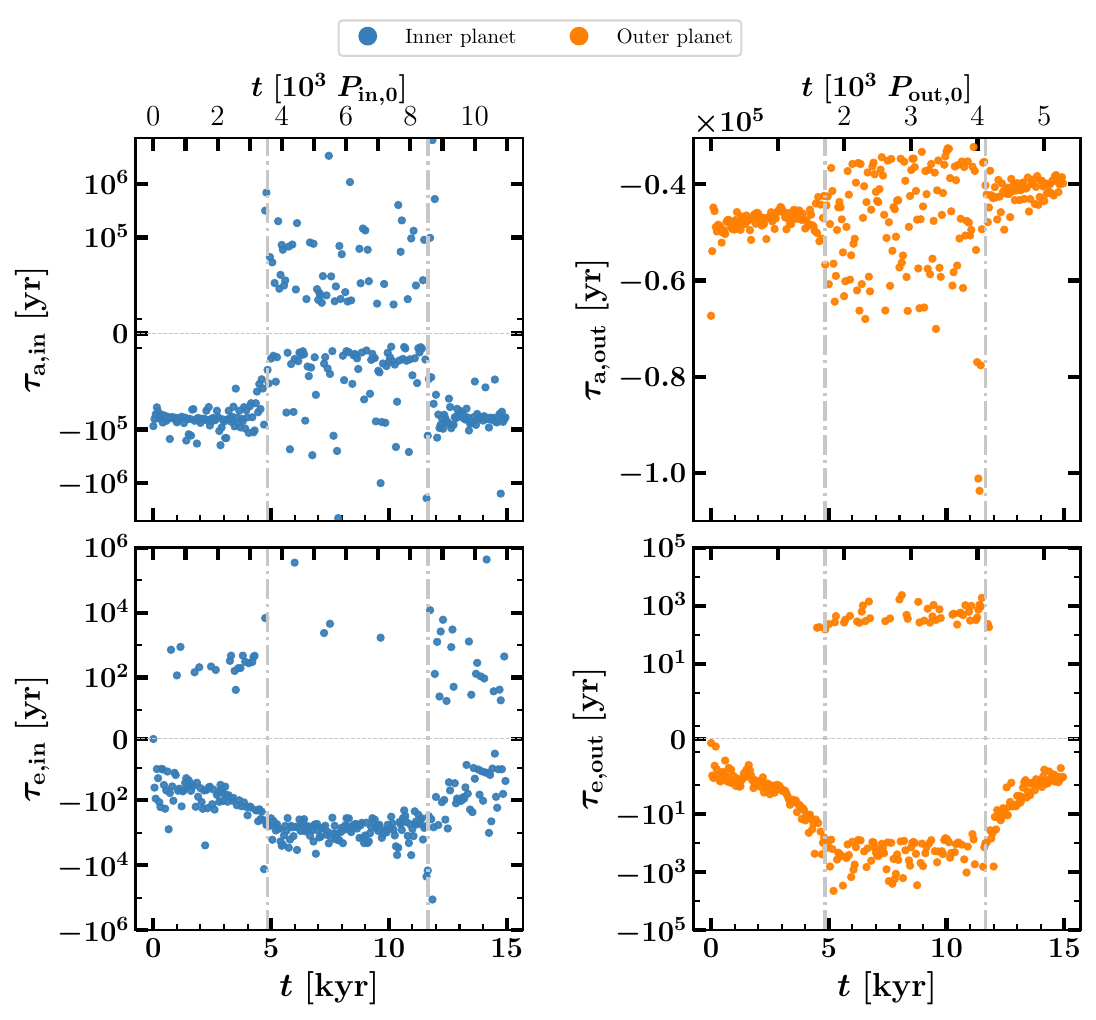}
	\caption{Temporal evolution of the orbital migration ($\tau_{a}$; top panels) and eccentricity damping ($\tau_{e}$; bottom panels) timescales for the inner (blue) and outer (orange) planets. 
	The lower horizontal axes show time in kyr, while the upper horizontal axes show time in units of $10^3$ initial orbital periods, $P_{\mathrm{in},0}$ and $P_{\mathrm{out},0}$, for the inner and outer planets, respectively.
		None of the data points reach zero; the minimum absolute values of the eccentricity damping timescale are 0.097 and 0.202 for the inner and outer planets, respectively.
		Vertical dot-dashed lines mark the times of 2:1 mean-motion resonance capture and escape, respectively.
		}
	\label{fig:tau_ae_column}
\end{figure}
%=====================================================================================================

For data visualisation, we employed a dual-scale plotting approach, using a symmetric logarithmic vertical axis and a linear horizontal axis. This scale was applied to the migration and eccentricity damping timescales of the inner planet, as well as to the eccentricity damping timescale of the outer planet. For the migration timescale of the outer planet, a linear--linear scale was adopted for all models, as no sign reversals were observed (except for the $h_0=0.03$ and $h_0=0.04$ models, which are non--resonant).
%For the migration timescale of the outer planet, we used a linear--linear scale in all models, since no sign reversals occurred in this case.
Figure \ref{fig:tau_ae_column} demonstrates how this method effectively captures both the wide dynamic range of the timescale magnitudes and the significant distinction between positive and negative values.
Figure~\ref{fig:tau_ae_column} shows the temporal evolution of the migration and eccentricity damping timescales, including the occurrence of sign reversals. The occurrence and physical origin of these sign reversals are discussed in Secs.~\ref{sec: results}, ~\ref{subsec: Bifurcation timescale}, and App.~\ref{appendix: extra simulations} (see especially Fig.~\ref{fig:snap_highreso_during_in_out}, with particular attention to the panels showing $a_{in}, a_{out}, \tau_{a,in}$, and $\tau_{a, out}$).
	
After extracting the migration and eccentricity damping timescales from the hydrodynamical simulations, we generalised the relation $\tau_e \approx h^2 \tau_a$ \citep{2014AJ...GS} by introducing a dimensionless factor, $\xi$, through the modified expression $\tau_e = \xi h^2 \tau_a$. Substituting this relation into Equation~\ref{eq:timescales} yields the explicit form of $\xi$ as,
%===========================================Equation: xi===============================================
\begin{equation}
\xi = \frac{\tau_{e}}{h^2\tau_{a}} %= \frac{-\frac{e^2}{(1-e^2)h^2}}{\tau_a \frac{\Gamma}{L} - \frac{1}{2}}.
\label{eq:xi_def}
\end{equation}
%=======================================================================================================
If our hydrodynamical results follow the relation, $\tau_e \approx h^2 \tau_a$, then $\xi$ should approach unity.
In the section~\ref{subsec: xi}, we evaluate the behaviour of $\xi$ in our simulations.

To examine the planet's response beyond the Type-I migration, we normalised the migration timescale by a reference value, $\tau_{a,0}$, that factors out the explicit dependencies on planet mass, disc surface density, and aspect ratio,  and allows us to compare the evolution of the migration timescales on a common basis. To derive this reference value, we set the eccentricity to zero in the eccentricity damping timescale equation and substituted $\Gamma$ with $\Gamma_0 = (q/h(r))^2 \Sigma(r) a^4 \Omega(r)^2$, where $\Omega(r)$ is the Keplerian angular velocity, $q$ is the planet to star mass ratio, and $\Sigma(r)$ is the disc surface density. The reference value was then obtained as follows,
%scaled migration timescale was then obtained as follows,
%=================================Equation: scaled migration timescale=================================
\begin{equation}
\tau_{a,0} = \frac{h_0^2}{2 q \Sigma_0} a,
\label{eq:scaled migration}
\end{equation}
%=======================================================================================================
where $\Sigma_0$ and $h_0$ are surface density and aspect ratio at the reference radius \(a_0 = 1\), respectively.
Here, $\Sigma_0$ is expressed in units of $M_\star/a_0^2$, and the equation is written in the code units adopted in our simulations, with $G = M_\star = a_0 = 1$.

\begin{figure} 
	\includegraphics[width=\columnwidth]{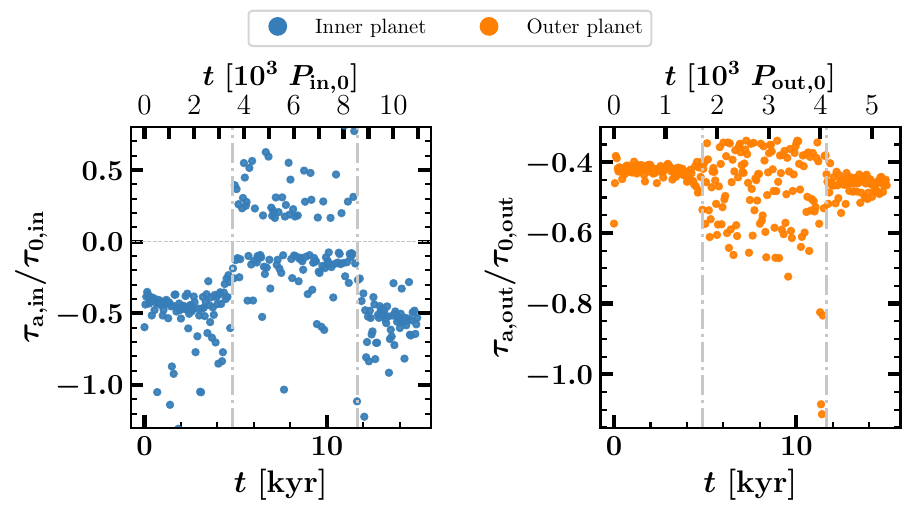}
	\caption{Temporal evolution of the migration timescale, scaled to $\tau_{a,0}$.}
	\label{fig: tau_a_scaled}
\end{figure}

Figure~\ref{fig: tau_a_scaled} presents the scaled migration timescales, $\tau_a/\tau_{a,0}$, of the inner and outer planets in the fiducial model. 
To reduce the variability associated with the eccentric phases, we focus on the intervals before and after the resonance, where the eccentricities are small. The evolution of the scaled migration timescale in these intervals can reflect changes in the planet-disc interaction as the planets migrate through the disc. 
In particular, the gas distribution around the planets can evolve through gas depletion and gap formation, while spiral density waves can produce additional radial structures. Especially in the low-viscosity regime considered here a single planet is able to open several gaps~\citep{dong2017ApJ...843..127D, dong2018multiple}. The perturbations generated by the two planets may also overlap, further modifying the local disc structure and the torques acting on the planets. Consequently, the evolution of the disc torque and power can reflect a combination of these effects, resulting in an evolving migration response and corresponding changes in the scaled migration timescale.
%Consequently, the evolution of the disc torque and power can reflect a combination of these effects, leading to changes in the migration timescale and its scaled value.}
	
%=======================================Figure: tau_a_scaled===========================================
% figure: tau_a_scaled
%====================================================================================================== 

\section{Results} \label{sec: results}
This section presents the results on how disc properties influence planetary dynamics. We show the migration, scaled migration, and eccentricity damping timescales for models with varying viscosity, surface density, and aspect ratio, building on the parameter study from our previous work (Af24). 

%====================================SubSection: Viscosity models================================
%======================================================================================================
\subsection{Viscosity models} \label{subsec: viscosity}
To examine the effect of disc viscosity on the timescales, we varied the viscosity parameter from the fiducial value of $\alpha_{\nu} = 10^{-5}$ to $\alpha_{\nu} = 10^{-6}, 10^{-4},$ and $10^{-3}$. We showed in Af24~(see their Fig.~5) that all viscous models were overstable except for the case with $\alpha_{\nu} = 10^{-3}$, which exhibited limit cycle behaviour.

\begin{figure*} 
	\centering
	
	\includegraphics[width=\textwidth]{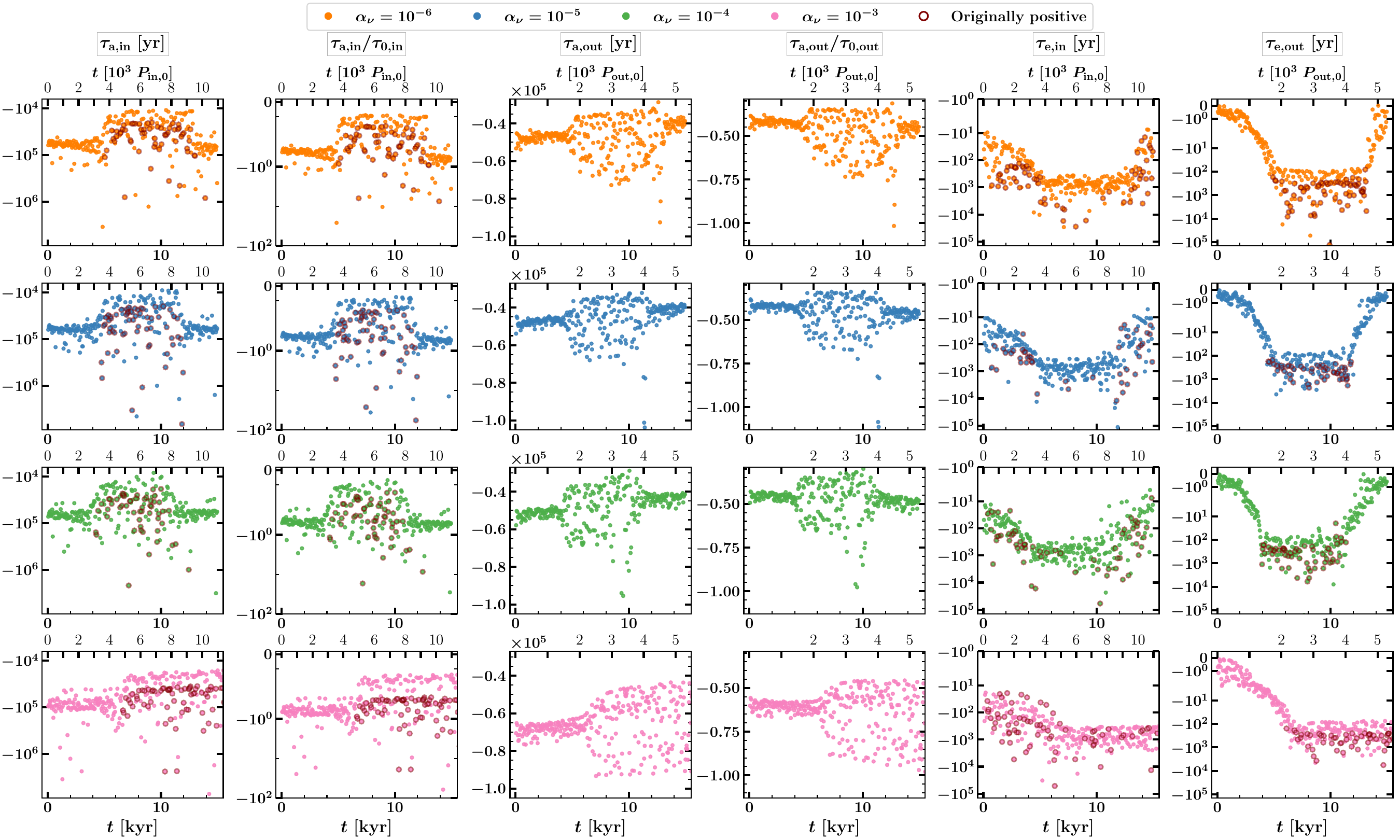}
	\caption{Time evolution of the migration timescale, scaled migration timescale, and eccentricity damping timescale for different viscosity models. The first row corresponds to $\alpha_{\nu} = 10^{-6}$ (orange), the second row to $\alpha_{\nu} = 10^{-5}$ (blue, fiducial model), the third row to $\alpha_{\nu} = 10^{-4}$ (green), and the last row to $\alpha_{\nu} = 10^{-3}$ (pink). The brown empty circles represent the positive values of timescales. For clarity in the compact multi-panel layout, the vertical-axis labels are indicated by the boxed titles at the top of each column rather than being repeated alongside individual panels. }
	\label{fig: alphas}
\end{figure*}

Figure~\ref{fig: alphas} shows the migration timescale, scaled migration timescale, and eccentricity damping timescale of the inner and outer planets for $\alpha_{\nu} = 10^{-6}, 10^{-5}, 10^{-4},$ and $10^{-3}$ (from top to bottom). The brown empty circles denote data points that were originally positive; for clarity, and to avoid leaving empty space between positive and negative values, and ease the comparison between the two branches of timescales, we flip the sign of the positive timescales and move them to the negative side of the axis.

%======================================== figure: alphas===============================================
%=====================================================================================================

The migration timescales of the inner and outer planet (columns 1 and 3) indicate that, at the beginning of the evolution, higher viscosity corresponds to longer migration timescales. In other words, the model with the highest viscosity exhibits the slowest migration (see Af24, Fig. 5). 
This behaviour arises because larger viscosity prevents the positive corotation torque from saturating, thereby reducing the net negative Lindblad torque and slowing down the migration~(see Af24, Sec. 4.2). 
On average, across all models, the inner planet has a migration timescale of approximately $10^4-10^5$~yr, while that of the outer planet is about $(3-8) \times 10^4$~yr.

To examine the evolution of the migration response across the viscous models, we compare the scaled migration timescales (columns 2 and 4) before and after resonance. Because the model with  $\alpha_{\nu} = 10^{-3}$ exhibits limit-cycle behaviour with the planets remaining in resonance, we do not include it in our timescale comparison here. 
For the inner planet, all three overstable models display a slight increase in the scaled migration timescale after resonance compared to before. 

The magnitude of this change increases towards lower viscosity. This trend is consistent with the stronger modification of the disc structure in the lower-viscosity models, including the formation of deeper partial gaps by the planets (see Fig.~\ref{fig:azi_combined}). Previous studies have shown that planets embedded in low-viscosity discs can open deeper partial gaps \citep{2006crida, 2012duffell}. In addition, in the low-viscosity regime considered here, the dissipation of planetary spiral waves can produce multiple radial structures in the gas distribution \citep{dong2017ApJ...843..127D, dong2018multiple}. The change in the scaled migration timescale can therefore reflect the combined effects of gap formation, spiral-wave structures, gas depletion, and other changes in the planet-disc interaction.

At the first glance, one might conclude that the scaled migration timescale difference of the outer planet in the $\alpha_{\nu} = 10^{-4}$ model is larger than in the less viscous model. However, as overstability happens earlier in this model, the system is more evolved after that and the planets are closer together at the end of the simulation that can affect the migration rate of the outer planet. If we compare the first $2$ kyrs after the resonance escape of all models, the results are consistent.
Across all viscous models, the magnitude of scaled migration timescale of the inner planet ($\nicefrac{\tau_{\rm a,in}}{\tau_{\rm 0,in}}$) remains within roughly $0.06$ to $1$.
	
The eccentricity damping is operated when a non-zero eccentricity exists. During the resonance time, when the planets attain their equilibrium eccentricities, $\tau_e$ of the inner planet is of the order of $10^2$ - $10^3$ yr for all viscous models, whereas that of the outer planet lies between $10$ and $10^3$~yr. Before resonance, $\tau_e$ gradually increases, while after resonance it decreases. This trend reflects the dependence of $\tau_e$ on the eccentricity: as the eccentricity grows or decays,
the damping timescale changes accordingly. Nevertheless, because the eccentricities remain very small outside the resonance phase, the corresponding damping timescale has no physical importance, as the planets have almost no eccentricity to be damped.	

%====================================SubSection: Surface density models================================
%======================================================================================================
\subsection{Surface density models} \label{subsec: surface density}
To examine the effect of disc surface density on the timescales, we varied the surface density at unit length from the fiducial value of $\Sigma_{0} = 1.13 \times 10^{-4} M_{\star}/r_0^2$ to $\Sigma_{0}/4$, $\Sigma_{0}/3$, $\Sigma_{0}/2$, $2\Sigma_{0}$, and $3\Sigma_{0}$. From our previous work~(see Figs.~8 and 9 in Af24), we expect that all models with various surface density values were overstable except for the case with $3\Sigma_{0}$, which was not overstable but instead captured in a 2:1 resonance.

\begin{figure*} 
	\centering
	
	\includegraphics[width=\textwidth]{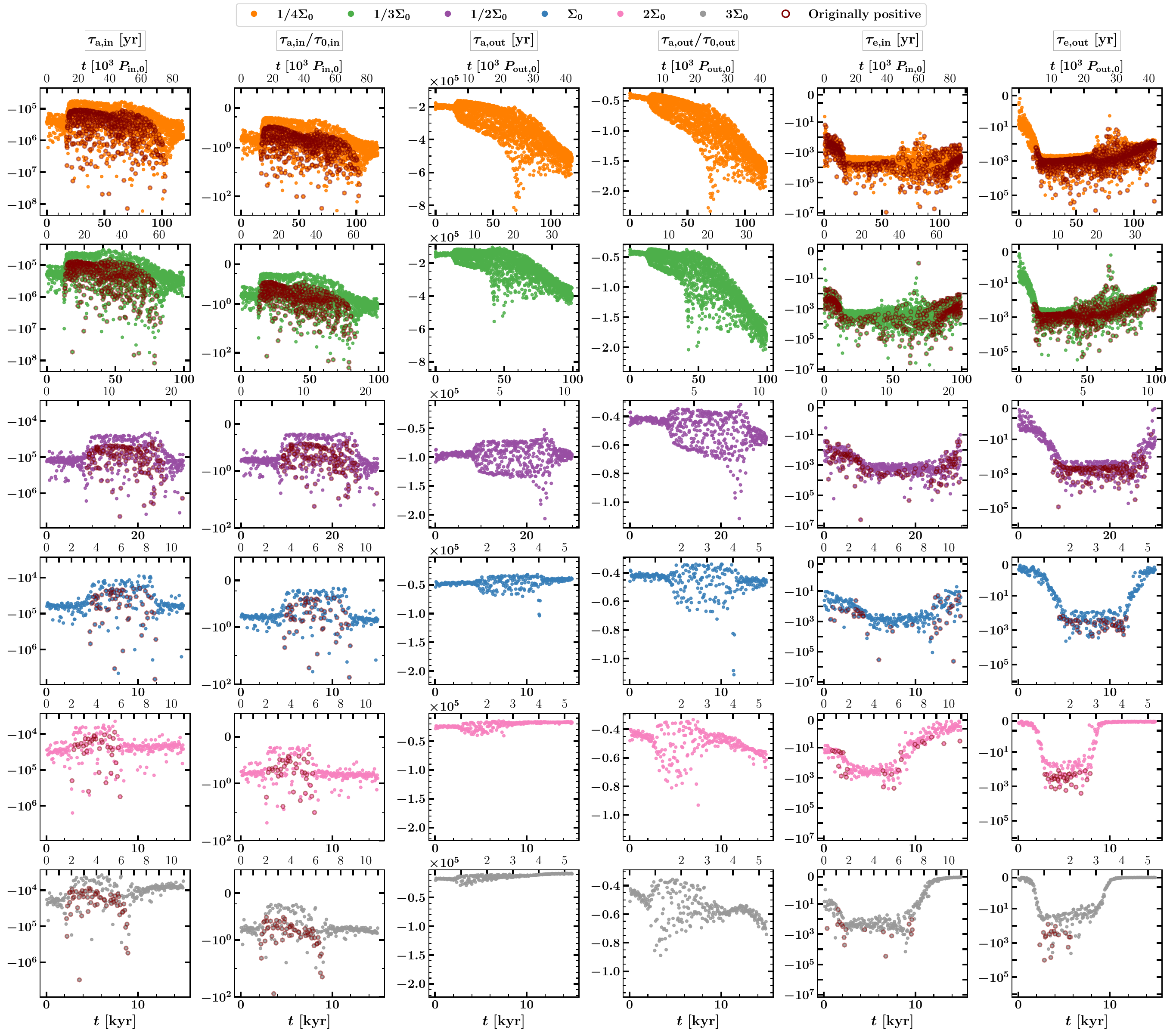}
	\caption{Time evolution of the migration timescale, scaled migration timescale, and eccentricity damping timescale for different surface density models. From the first to the last row, the cases correspond to $\tfrac{1}{4}\Sigma_0$ (orange), $\tfrac{1}{3}\Sigma_0$ (green), $\tfrac{1}{2}\Sigma_0$ (purple), $\Sigma_0$ (blue, fiducial model), $2\Sigma_0$ (pink), and $3\Sigma_0$ (gray), respectively. The brown empty circles represent the positive values of timescales. Please note that the vertical-axis labels are indicated by the boxed titles at the top of each column.
	}
	\label{fig: sigmas}
\end{figure*}

Figure~\ref{fig: sigmas} presents the migration timescale, scaled migration timescale, and eccentricity damping timescale of the inner and outer planets for $\Sigma_{0}/4$, $\Sigma_{0}/3$, $\Sigma_{0}/2$, $2\Sigma_{0}$, and $3\Sigma_{0}$ (from top to bottom, respectively). 

%======================================= figure: sigmas================================================
%=====================================================================================================

The migration timescales in columns 1 and 3 demonstrate that, early in the evolution, planets in higher surface density discs migrate more rapidly, as evidenced by their shorter migration timescales. This behaviour aligns with the evolution of their semi-major axes (Figs. 8 and 9 of Af24). The faster migration in these models results from the stronger net torque exerted by the more massive disc. For the inner planet, the migration timescale spans from roughly $10^4 $~yr to about $10^7 $~yr. For the outer planet, the migration timescale ranges from $(1-7) \times10^5$~yr.

To analyse the scaled migration timescale (columns 2 and 4), we consider only the values before and after the resonant phase, as described in Sec.~\ref{sec:methods}. For both planets and across all models, the scaled migration timescale drops after the resonance. This evolution reflects changes in the planet-disc interaction as the planets migrate through the disc. 

For the inner planet, the difference between the scaled migration timescales before and after the resonance increases with decreasing disc surface density in all models. 
This trend is consistent with the stronger evolution of the disc structure in the lower-density models, where the slower migration allows more time for the disc to respond to the planets and for partial gaps to develop. 
This also matches Fig.~\ref{fig:azi_combined}, which shows that the gap opened by the inner planet generally deepens as the surface density decreases. 
The only exception is the $3\Sigma_{0}$ model, in which the inner planet opens a slightly deeper gap than in the $2\Sigma_{0}$ model. This behaviour can be attributed to the smaller orbital radius of the inner planet in the $3\Sigma_{0}$ model, where it experiences a lower effective viscosity ($\sim 0.35$ times between $ r = 0.7$ and $2$) and smaller aspect ratio($\sim 0.75$ times between $ r = 0.7$ and $2$), allowing a deeper gap to form~\citep{2006crida}.

For the outer planet, the $\Sigma_{0}/4$, $\Sigma_{0}/3$, $\Sigma_{0}/2$, and $\Sigma_{0}$ models show a similar trend to the inner planet, with larger changes in the scaled migration timescale at lower surface densities. This trend is consistent with the deeper partial gaps seen in the lower-density models (Fig.~\ref{fig:azi_combined}).
However, in the $2\Sigma_{0}$ and $3\Sigma_{0}$ models, the change in the scaled migration timescale appears larger than in the $\Sigma_{0}$ and $\Sigma_{0}/2$ models.
It is important to note that in these two models, the planets capture in the resonance and escape from it earlier. 
As a result, the post-resonance evolution is sampled over a longer time interval (approximately 5-7~kyr) than in other models, which cover only up to about 3~kyr after resonance. In the $2\Sigma_{0}$ model, up to about 3~kyr after leaving resonance (corresponding to approximately 11~kyr), the change in the scaled migration timescale remains comparable to that in the $\Sigma_{0}$ model and is consistent with the gap depth formed in this case. 

As mentioned in Sec.~\ref{subsec: viscosity}, the eccentricity damping timescale is meaningful only during the resonance phase. During resonance, the eccentricity damping timescale for the inner planet ranges from approximately $10^3$ to  $10^5$~yr in the model with the lowest surface density ($\Sigma_0/4$), and from about $10$ to $10^3$~yr in the model with the highest surface density ($3 \Sigma_0$). For the outer planet, it varies from roughly $10^2$ to $10^4$~yr in the lowest surface density model, and from about $10$ to $10^2$~yr in the highest surface density model.

%====================================SubSection: Aspect ratio models================================
%======================================================================================================
\subsection{Aspect ratio models} \label{subsec: aspect ratio}
To examine the effect of the disc aspect ratio on the timescales, we varied the aspect ratio from the fiducial value of $h_{0} = 0.05$ to $h_{0} = 0.03$, $h_{0} = 0.04$, and $h_{0} = 0.06$. As the results of models with different $h_0$s show in Af24~(see their Figs.~10 and 11), only the model with $h_{0} = 0.06$ was overstable, while the other two models were neither overstable nor resonant.

\begin{figure*} 
	\centering

	\includegraphics[width=\textwidth]{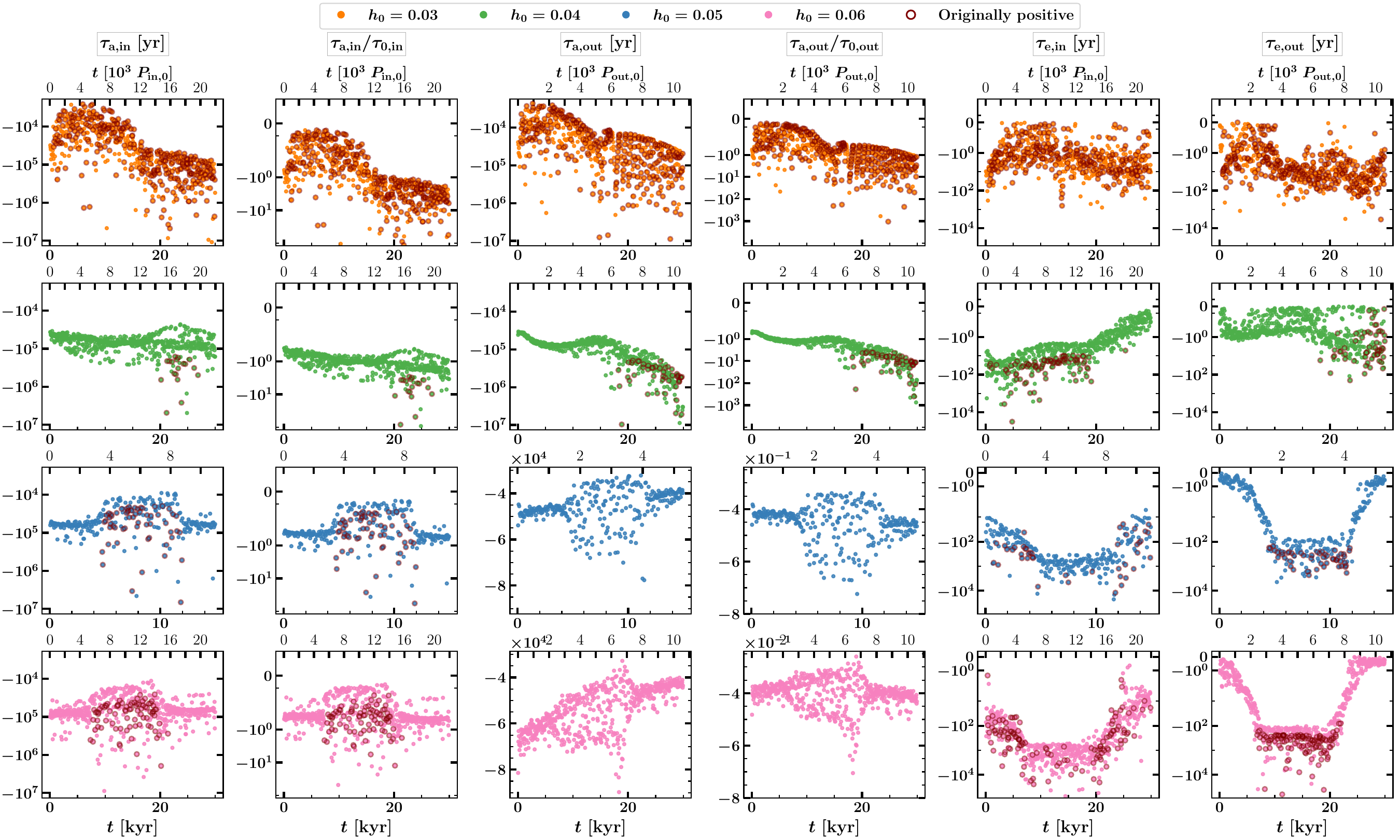}
	\caption{Time evolution of the migration timescale, scaled migration timescale, and eccentricity damping timescale for different aspect ratio models. The first row corresponds to $h_{0} = 0.03$ (orange), the second row to $h_{0} = 0.04$ (green), , the third row to $h_{0} = 0.05$ (blue, fiducial model) and the last row to $h_{0} = 0.06$ (pink). The brown empty circles represent the positive values of timescales. Please note that the vertical-axis labels are indicated by the boxed titles at the top of each column.
	}
	\label{fig: hs}
\end{figure*}

Figure~\ref{fig: hs} shows the migration timescale, scaled migration timescale, and eccentricity damping timescale of the inner and outer planets for $h_{0} = 0.03$, $h_{0} = 0.04$, $h_{0} = 0.05$, and $h_{0} = 0.06$ (from top to bottom, respectively). 
%The brown empty circles indicate data points that were originally positive; for clarity, and to avoid leaving empty space between positive and negative values, we shifted them to the negative axis.

%========================================== figure: hs=================================================
%=====================================================================================================

The migration timescales shown in columns 1 and 3 for the inner and outer planets, respectively, indicate that at the early stages of the evolution, the migration timescale increases with increasing disc aspect ratio, $h_0$. This implies faster migration in discs with smaller aspect ratios, consistent with expectations from torque relation, where the torque scales inversely with the square of the disc aspect ratio \citep[e.g.][]{2002ApJ...565.1257T}.
However, the subsequent evolution differs for the models with $h_0 = 0.03$ and $h_0 = 0.04$. In the $h_0 = 0.03$ model, the slope of the migration timescale increases sharply for both the inner and outer planets, indicating a significant slowdown of migration. 
%This behaviour is consistent with the evolution of the planets’ semi-major axes in this model (see AF24, Fig. 11). 
In the $h_0 = 0.04$ model, an increase in the migration timescale is also observed for both planets, although the slope is noticeably shallower for the inner planet than for the outer one. In this case, the inner planet migrates only slightly faster than in the fiducial model, while the migration of the outer planet is comparable to that in the fiducial model with $h_0 = 0.05$ (see AF24, Fig. 11).

In the $h_0 = 0.03$ model, a clear increase in the scaled migration timescale is observed for both planets, indicating a substantial change in the migration response. Figure \ref{fig:azi_combined} shows that the inner planet opens a gap with a depth of approximately 90 percent, while the outer planet opens an even deeper gap exceeding 90 percent. The deep gaps are consistent with the significantly reduced migration speeds observed in this model.

In the $h_0 = 0.04$ model, the scaled migration timescale also increases for both planets. As shown in Fig.~\ref{fig:azi_combined}, the inner planet opens a gap with a depth of about 60 percent, while the outer planet opens a gap with a depth of approximately 90 percent. These gap structures are consistent with the reduced migration speeds observed in this disc.

The $h_0 = 0.05$ and $h_0 = 0.06$ models show only a modest increase in the scaled migration timescale. The change is smaller in the $h_0 = 0.06$ model than in the $h_0 = 0.05$ case, consistent with Fig.~\ref{fig:azi_combined}, which shows that only partial gaps are formed in these discs, with the shallowest gaps occurring in the $h_0 = 0.06$ model.

As mentioned before, the eccentricity damping timescale is meaningful only if a planet's orbit is eccentric. Therefore, it is not defined for the models with $h_0 = 0.03$  and $h_0 = 0.04$, since the eccentricities remain approximately zero throughout the system evolution. In contrast, for the $h_0 = 0.05$  and $h_0 = 0.06$ models, during resonance, the eccentricity damping timescale of the inner planet ranges from $10^2$ to $10^4$~yr, while that of the outer planet varies from $10^2$ to $10^3$~yr.

%====================================Subsection: Behavior of $\xi$=====================================
%======================================================================================================

\subsection{Deviation from the $\tau_e \approx h^2 \tau_a$ relation} \label{subsec: xi}

In this section, we examine the validity of the commonly adopted relation $\tau_e = \xi h^2 \tau_a$ and investigate whether a constant value of $\xi$ provides an adequate description of the hydrodynamic evolution of resonant planetary systems.

\begin{figure} 
	
	\includegraphics[width=\columnwidth]{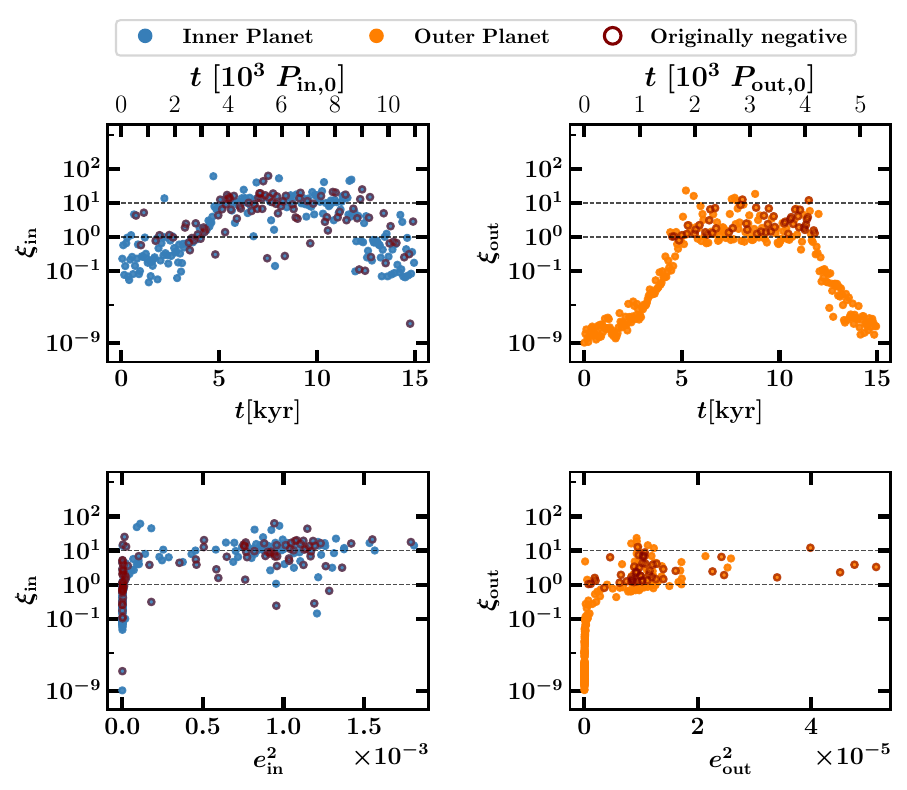}
	\caption{Variations of the parameter $\xi$ for the inner (blue) and outer (orange) planets. The first row shows the time evolution of $\xi$.
	The second row shows $\xi$ as a function of eccentricity, demonstrating that the ratio of the eccentricity damping to migration timescales does not depend on eccentricity. The brown empty circles represent the negative values of timescales.}
	\label{fig: xi}
\end{figure}

The upper panels of Fig.~\ref{fig: xi} show $\xi$ over time for our fiducial model. Before and after resonance, both planets exhibit relatively small values of $\xi$. During resonance, however, the amplitude of $\xi$ increases, and sign reversals are observed for both planets, originating from time reversal of migration and eccentricity damping timescales.

To quantify this variability, we calculated the distribution of $\xi$ within the resonant interval. For the inner planet, the median value of $|\xi|$ is $10.8$, with the central $68\%$ of the distribution lying between $4.6$ and $16.9$. The corresponding values for the outer planet are considerably smaller, with a median of $1.74$ and a central range of $0.84$ to $4.70$. These results demonstrate that $\xi$ changes approximately one order of magnitude during resonance and therefore cannot be represented by a single constant value.

The distributions also contain occasional extreme values. These rare excursions strongly bias the arithmetic mean, whereas the median remains robust.  For example, the mean value of $|\xi_{in}|$  is approximately $65.6$, while the corresponding median is $10.8$, demonstrating that the arithmetic mean is not representative of the typical value of $\xi$. For this reason, the median is adopted throughout this work as the characteristic value of $\xi$, rather than the arithmetic mean.

The lower panels of Fig.~\ref{fig: xi} show $\xi$ as a function of $e^2$. No clear dependence on eccentricity is observed, consistent with the weak Pearson correlation coefficients ($r = -0.1$ and $r = 0.01$ for the inner and outer planets, respectively).

%===========================================Figure: xi=================================================
% figure: xi
%====================================================================================================== 

\begin{figure} 
	\includegraphics[width=\columnwidth]{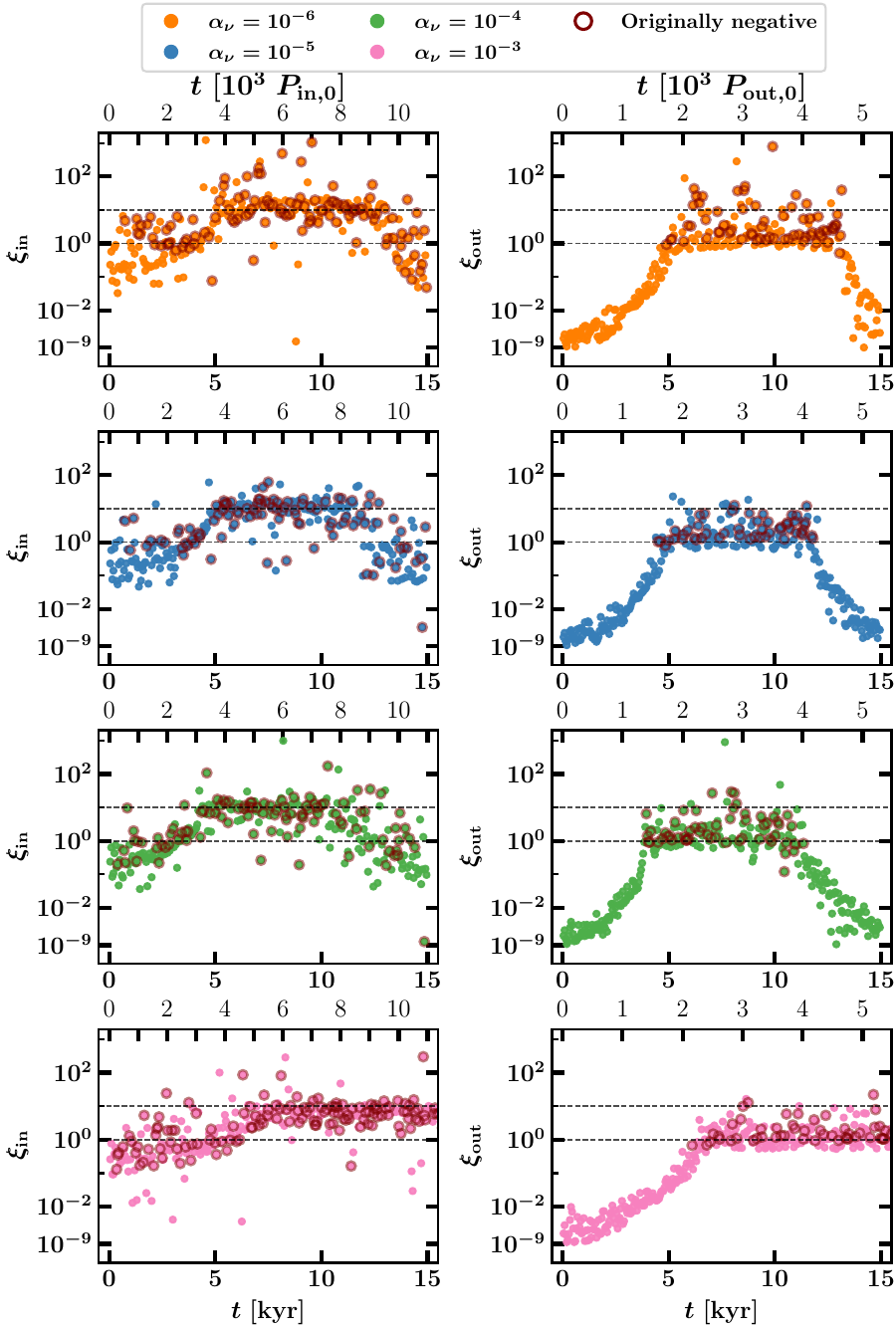}
	\caption{Variations of the parameter $\xi$ for different viscosity models, shown for the inner planet (left) and the outer planet (right). From top to bottom, the panels correspond to models with $\alpha_{\nu} = 10^{-6}, 10^{-5}, 10^{-4},$ and $10^{-3}$. The brown empty circles represent the negative values of timescales.
	}
	\label{fig:xi_alphas}
\end{figure}

\begin{figure} 
	\includegraphics[width=\columnwidth]{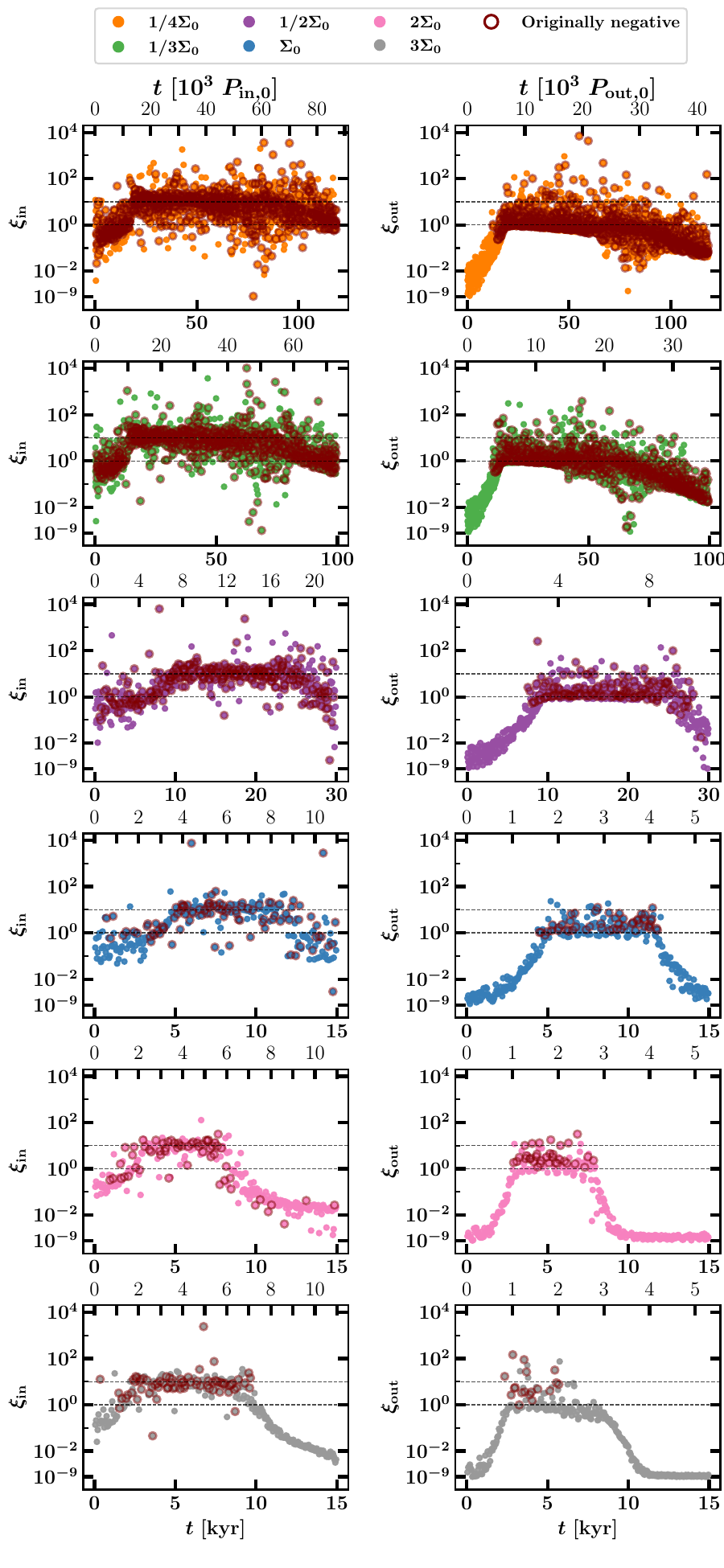}
	\caption{Variations of the parameter $\xi$ for different surface density models, shown for the inner planet (left) and the outer planet (right). From top to bottom, the panels correspond to models with $\Sigma_{0} = \Sigma_{0}/4, \Sigma_{0}/3, \Sigma_{0}/2, 2\Sigma_{0},$ and $3\Sigma_{0}$. The brown empty circles represent the negative values of timescales.
	}
	\label{fig:xi_sigmas}
\end{figure}

\begin{figure} 
	\includegraphics[width=\columnwidth]{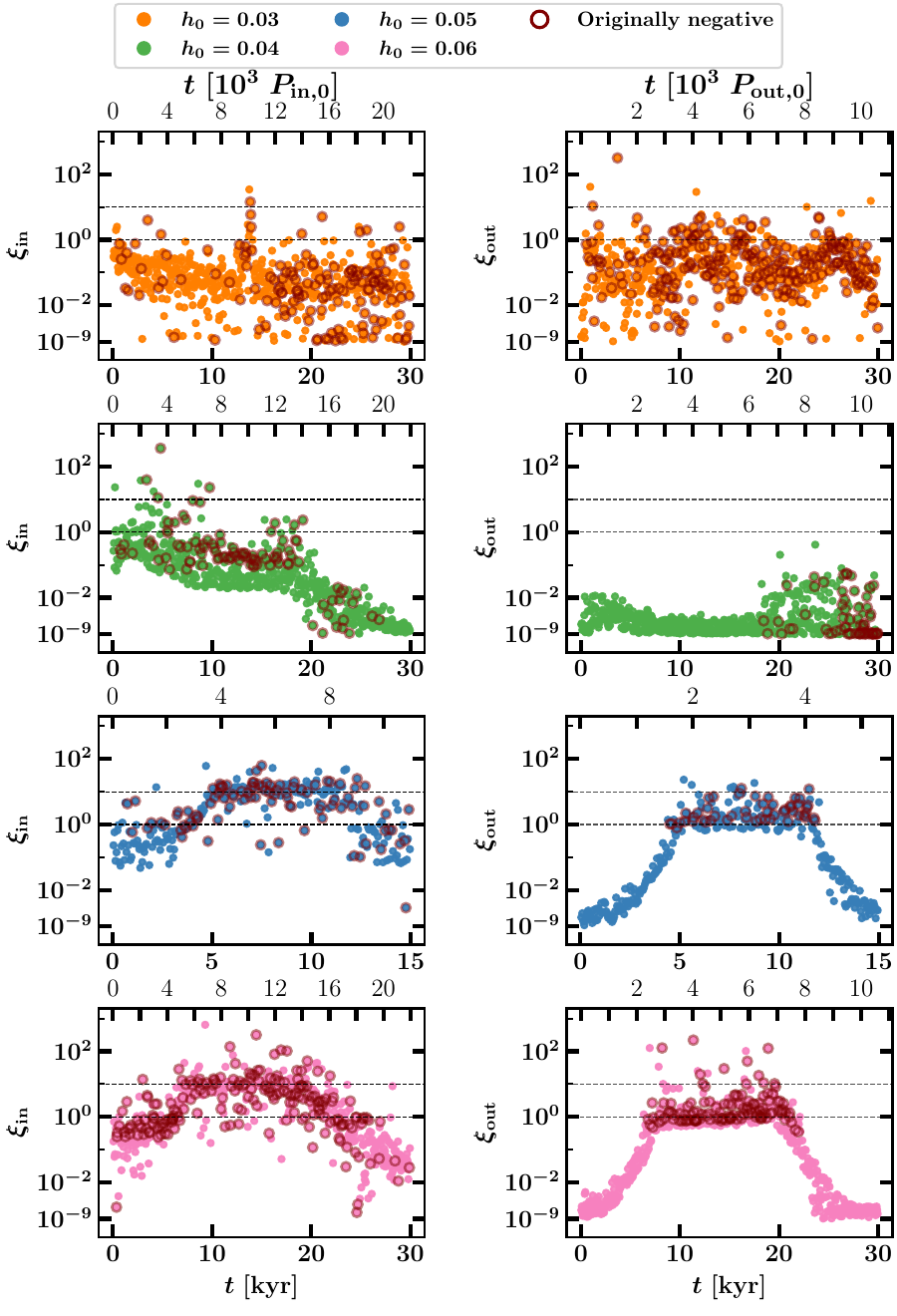}
	\caption{Variations of the parameter $\xi$ for different aspect ratio models, shown for the inner planet (left) and the outer planet (right). From top to bottom, the panels correspond to models with $h_{0} = 0.03, 0.04, 0.05,$ and $0.06$. The brown empty circles represent the negative values of timescales.
	}
	\label{fig:xi_hs}
\end{figure}

Figures~\ref{fig:xi_alphas}--\ref{fig:xi_hs} demonstrate that these conclusions are not restricted to the fiducial model. Across the explored viscosities, surface densities, and aspect ratios, all models undergoing resonance capture exhibit the same qualitative behaviour. $\xi$ remains strongly time dependent and spans a broad range of values throughout the resonant phase. By contrast, the two models that do not experience resonance capture ($h_0=0.03$ and $0.04$) display nearly constant and substantially smaller values of $\xi$. These results indicate that, although the characteristic values of $\xi$ depend on the adopted disc parameters, its pronounced temporal variability is a generic feature of resonant evolution. A statistical comparison of the distributions of $|\xi|$ for all models is presented in Appendix~\ref{app:violet_plot}.

Overall, these results show that the commonly adopted approximation of a constant $\xi$ does not accurately describe the hydrodynamic evolution of resonant planetary systems. Instead, $\xi$ should be regarded as a time-dependent quantity whose value evolves continuously throughout the resonant phase.

%=========================================Figure: xi_alphas============================================
%====================================================================================================== 

%=========================================Figure: xi_sigmas============================================
%====================================================================================================== 

%=========================================Figure: xi_hs================================================
%====================================================================================================== 

%====================================Section: N-body Simulations====================================
%======================================================================================================
\section{N-body simulations} \label{sec:nbody} 
In this section, we investigate several N-body methods for simulating two of our hydrodynamical models, the fiducial and $\nicefrac{1}{4}\Sigma_0$ models. We aim to examine if/how~much using different approaches for migration and eccentricity damping timescales result in various outputs. In all of our approaches, we use identical initial values for the disc properties (if needed), planets, and run-time as in the hydrodynamical models. For all of the N-body simulations, we used a version of modified \texttt{REBOUND} code \citep{2012A&A...537A.128R} employing \texttt{WHFAST} integrator \citep{2015MNRAS.452..376R} with a timestep of $10^{-3}\,$\si{\yr}. Only in two simulations, in which we applied the type-I migration of the \texttt{REBOUNDx} code \citep{2020MNRAS.491.2885T} to our setups, we used \texttt{IAS15} adaptive time~step integrator \citep{2015MNRAS.446.1424R} because \texttt{IAS15} is the default integrator in \texttt{REBOUNDx} while using \texttt{WHFAST} needs defining a new operator. As our focus is on overstability in the 2:1 resonance and no close encounter is expected for such systems, using a sufficiently small constant time~step with the \texttt{WHFAST} integrator would produce almost identical outputs to using the adaptive timestep integrator \texttt{IAS15} (see Appendix.~\ref{appendix: integrators}).

\subsection{N-body approaches} \label{nbody-apps}
Inspired by the results in Sec.~\ref{subsec: surface density}, and also the commonly used
methods in the N-body simulations of resonant systems, the following approaches are considered.
\begin{enumerate}
	\item \emph{Constant timescales}: The migration and eccentricity damping timescales are considered constant during the whole simulation. For every model, we ran 1000 simulations, each of them used a random set of $(\tau_{\rm a,in}, \tau_{\rm a,out}, \tau_{\rm e,in}, \tau_{\rm e,out})$. Distributions which the timescales are drawn are driven from the results of corresponding simulations in Fig.~\ref{fig: sigmas} and are shown in Fig.~\ref{fig:model1dists}. 
	
	From the drawn timescales, $\tau_{\rm L}$ is calculated in every timestep via
	\begin{equation} \label{eq: taul1}
	\frac{1}{\tau_{\rm L}} = \frac{1}{2 \tau_{\rm a}}-\frac{e^2}{(1-e^2)\tau_{\rm e}}.
	\end{equation}
	While $\tau_{\rm e}$ is constant during the simulation, $\tau_{\rm L}$ is changed as the planet acquires eccentricity. The timescales $\tau_{\rm e}$  and $\tau_{\rm L}$ modify the acceleration of the planet by the relation in \citet{2000MNRAS...PL} that reads
	\begin{equation} \label{eq: paplar2000}
	\vec{a} = -\frac{\vec{v}}{\tau_{\rm L}} - 2 \frac{\vec{r}.\vec{v}}{r^2 \tau_{\rm e}} \vec{r},
	\end{equation}
	where $\vec{v}$ and $\vec{r}$ are velocity and position vector of the planet. The planets are initialized with no eccentricities.

\begin{figure} 
	\includegraphics[width=\columnwidth]{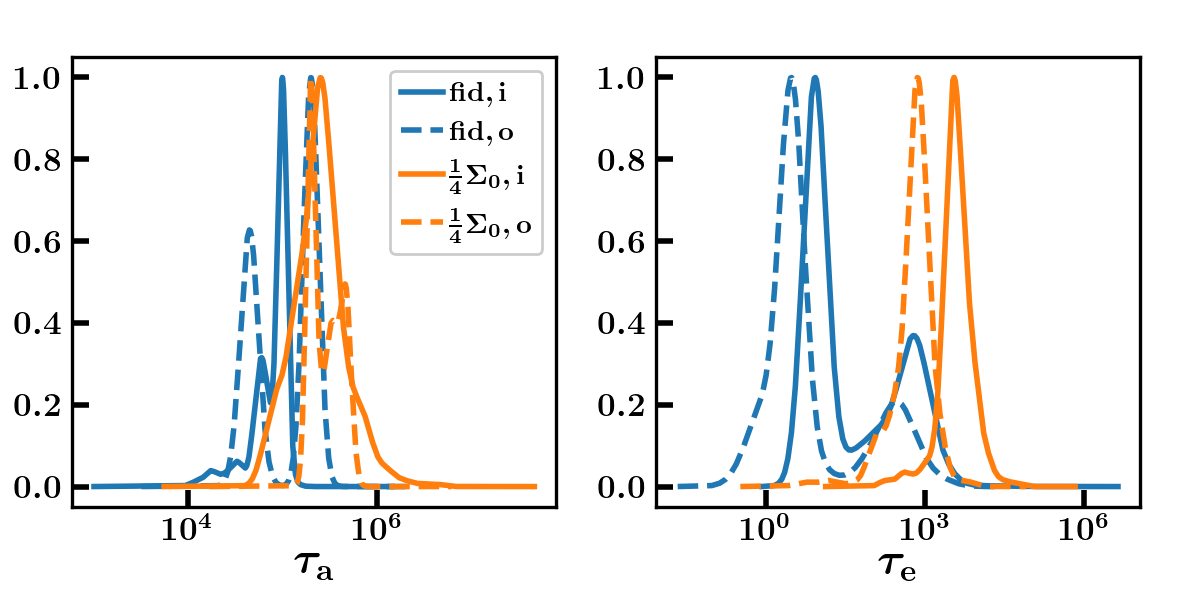}
	\caption{The distributions of migration and eccentricity damping timescales taken from the results of the two examined hydrodynamical models in the N-body simulations.}
	\label{fig:model1dists}
\end{figure}

\item \emph{Scaled timescales}: We initially choose a set of scaled migration timescale $\tau_{\rm a}/\tau_{0}$ and $\xi$ for each planet and keep them constant during the simulation. After a scaled migration timescale is chosen from its distribution, it is multiplied by $\tau_{\rm a,0}$ for each planet at every timestep. Because the planet's semi-major axis is changing by time,$\tau_{\rm a,0}$ should be updated accordingly. Then, a random $\xi$ is drawn and $\tau_{\rm e} = \xi h^2 \tau_{\rm a}$ is computed. These quantities are used by Eqs.~\ref{eq: taul1} and \ref{eq: paplar2000} to apply the effect of migration on the planet. The used distributions are given in Fig.~\ref{fig:model2dists}. For each model, 1000 simulations are run.

\begin{figure} 
	\includegraphics[width=\columnwidth]{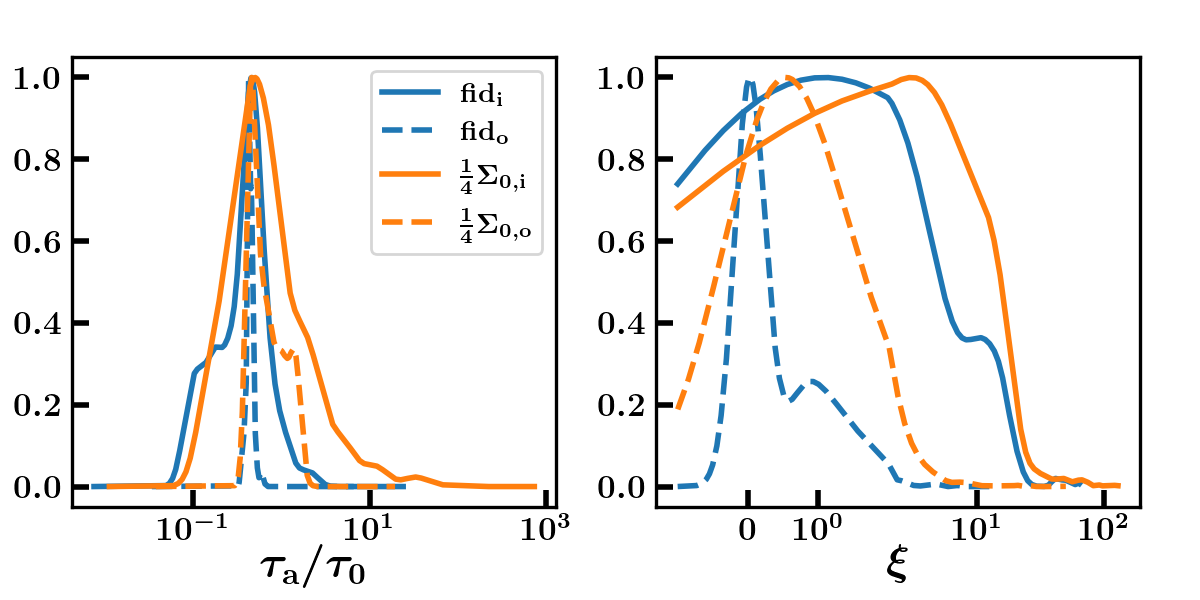}
	\caption{The distributions of scaled migration timescales and $\xi$s used in approach~(ii) that are taken from the results of the two examined hydrodynamical models.}
	\label{fig:model2dists}
\end{figure}

\item \emph{Scaled $\tau_{\rm e}$}: In this approach, first $\tau_{\rm L}=L/\Gamma$ is calculated, where the torque on the planet $\Gamma$ is given by the formula in \cite{2010MNRAS.401.1950P} and \cite{2011MNRAS.410..293P}. For this approach, we need to give the disc properties including surface density and aspect ratio profiles to the code. Then, the migration and eccentricity damping timescales are determined by
\begin{align} \label{eq: taul2}
&\tau_{\rm a} = \tau_{\rm L} \left(\frac{1}{2}-\frac{e^2}{(1-e^2)(\tau_{\rm e}/\tau_{\rm a})} \right) = \tau_{\rm L} \left(\frac{1}{2}-\frac{e^2}{(1-e^2)\xi h^2} \right)\\
& \tau_{\rm e} = \xi h^2 \tau_{\rm a}.
\end{align}
In the first timestep, $\tau_{\rm e}$ and eccentricity are zero but they are updated afterwards. This means that, $\tau_{\rm a}$ is calculated using the torque in that time~step and $\tau_{\rm e}$ in the previous one. $\xi$, that is constant during each simulation, is a random number from its corresponding distribution in the side panels in Fig.~\ref{fig:nbody_model3}. Identical to the previous approaches, these timescales modify the acceleration of the planet in each time~step. Similar to the two previous approaches, 1000 simulations per model are run.

\item \emph{Torque calculated timescales}: We give the disc and planet properties to the N-body code and the timescales are calculated using the torque on the planets in the simulations with a modified version of the \texttt{REBOUND} code as explained in \cite{2021A&A...AK}. We also added \cite{2018Kanagawa} scheme to the code in order to include migration with gap-opening. In another set of simulations, we experiment the type-I migration of the \texttt{REBOUNDx} code too. For these cases, we only give the disc surface density and aspect ratio profiles beside the planets initial semi-major axes.

\end{enumerate}

%======================================= Nbody results==================================================
\subsection{Results of N-body simulations} \label{sec:nbody-results}
%========================================= nbody-model 1 ===============================================
The characteristic of overstability is a librating resonance angle with growing amplitude that results into breaking the resonance and circulating the resonance angle. To find the overstable systems~(OS), we go through the following procedure. First, we find when  the system's orbital period ratio stays between \num{1.99} and \num{2.01}. Then, we calculate the resonance angle for the 2:1 MMR in that time span. Thenceforth, we find the local minima and maxima in the resonance angle and monitor if the minima is descending and maxima is ascending. Finally, we check whether the planets eccentricities are damped to zero with an absolute  tolerance of \num{0.001}. The final stage guarantees that the resonance is broken at the end of simulation and the planet's eccentricities are damped.

\begin{figure} 
	\includegraphics[width=\columnwidth]{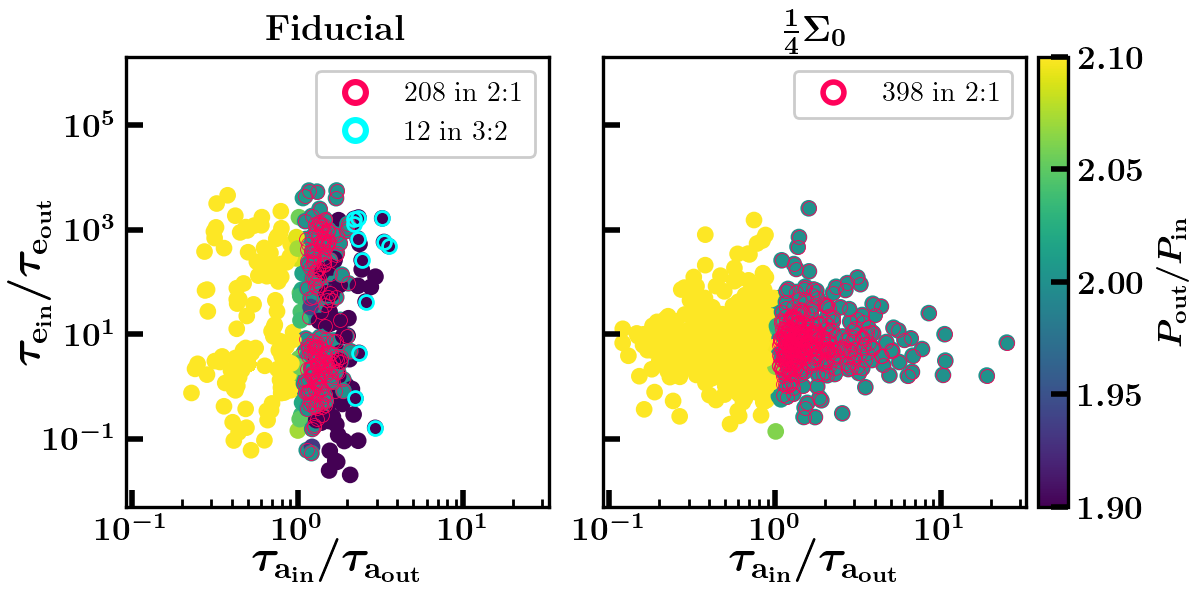}
	\caption{Orbital period ratio of the outer to the inner planet for the simulations in approach~(i). The left panel shows the results for the fiducial model and the right panel for the $\nicefrac{1}{4}\Sigma_0$ model. The horizontal/vertical axes correspond to the migration/e-damping timescale ratios of the inner to the outer planet. None of these models show overstable behaviour but many ended in various resonances as marked in the plots.    }
	\label{fig:nbody_model1}
\end{figure}

None of the \num{2000} simulations in the approach~(i) with constant timescales show overstable behaviour. 220 systems ($\sim 20\%$) in the fiducial model and 398 systems ($\sim 40\%$) in the model with $\nicefrac{1}{4}\Sigma_0$ remain in a resonance until end of the simulations. The larger percent of resonant systems in the $\nicefrac{1}{4}\Sigma_0$ is because the migration timescales are larger than the fiducial model. Therefore, after the planets reach a resonance, they stay in it and migrate together. As Fig.~\ref{fig:nbody_model1} shows, the resonant systems in both models stay in 2:1 resonance, except 12 systems~($\sim 1\%$) in the fiducial model that are in 3:2. The models with slightly faster migration for the inner planet, shown in yellow, went through diverging migration. Regardless of the eccentricity damping timescale ratios, there is a threshold on migration timescale ratios for the final configuration of the system. Our systems with $1 \lesssim \nicefrac{\tau_{\rm a,in}}{\tau_{\rm a,out}} \lesssim 2$ in the fiducial model, and $\nicefrac{\tau_{\rm a,in}}{\tau_{\rm a,out}} > 1$ in the $\nicefrac{1}{4}\Sigma_0$ model result in 2:1 resonance. It indicates that in addition to the ratio of the migration timescales, the value of the timescales , which determine how fast our planets migrate, can affect the final configuration.

%===================================nbody-model 4: approach 2======================================================
\begin{figure} 
	\includegraphics[width=\columnwidth]{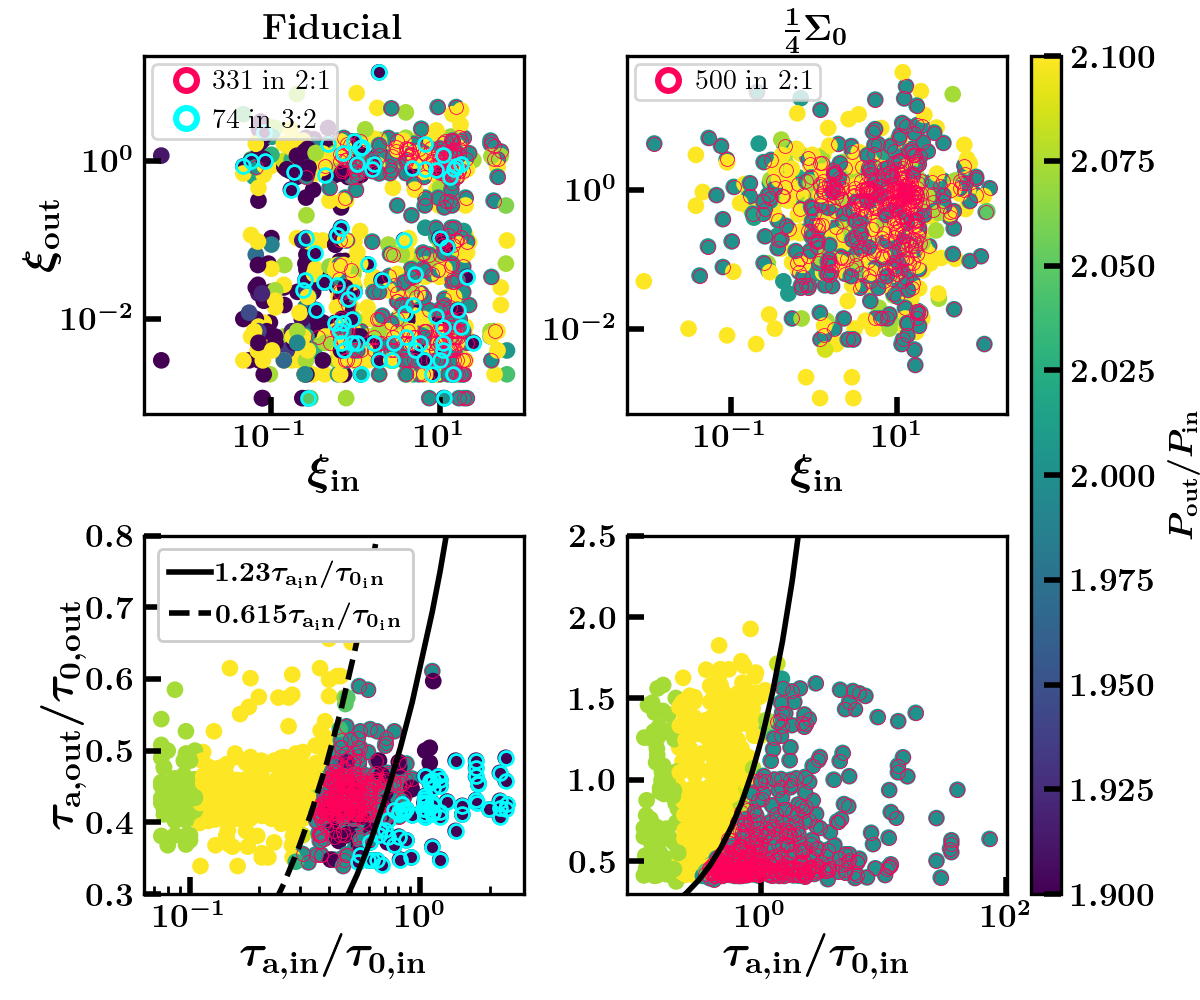}
	\caption{The top panels show the $\xi$ values for the scaled timescales approach and the bottom ones demonstrate the scaled migration timescales. The solid and dashed lines in the bottom panels respectively represent the criteria $\nicefrac{\tau_{\rm a,in}}{\tau_{\rm a,out}}>1$ and  $\nicefrac{\tau_{\rm a,in}}{\tau_{\rm a,out}}<2$.}
	\label{fig:nbody_model4}
\end{figure}

The results of approach~(ii) with scaled timescales, are shown in Fig.~\ref{fig:nbody_model4}. No system is overstable in these \num{2000} simulations and about half of them, \num{405}(~$40\%$) in the fiducial and \num{500}(~$= 50\%$) in the $\nicefrac{1}{4}\Sigma_0$ model, stay in a resonance until end of the simulation. The results of this approach are in agreement with approach~(i). The solid and dashed lines in the bottom panels of Fig.~\ref{fig:nbody_model4} represent the criteria $\nicefrac{\tau_{\rm a,in}}{\tau_{\rm a,out}}>1$ and  $\nicefrac{\tau_{\rm a,in}}{\tau_{\rm a,out}}<2$. Having $\nicefrac{q_{\rm out}}{q_{\rm in}}=2$ and $\nicefrac{a_{\rm out}}{a_{\rm in}} = \nicefrac{2}{1.23}$, our criteria read
\begin{align} \label{eq: tratio_model2}
&1<\frac{\tau_{\rm a,in}}{\tau_{\rm a,out}} <2,\\
&\frac{\tau_{\rm 0,out}}{\tau_{\rm 0,in}}<\frac{\tau_{\rm a,in}/\tau_{\rm 0,in}}{\tau_{\rm a,out}/\tau_{\rm 0,out}} <2\frac{\tau_{\rm 0,out}}{\tau_{\rm 0,in}},\\
&\frac{1}{1.23}<\frac{\tau_{\rm a,in}/\tau_{\rm 0,in}}{\tau_{\rm a,out}/\tau_{\rm 0,out}} <\frac{2}{1.23},\\
& 0.615 \frac{\tau_{\rm a,in}}{\tau_{\rm 0,in}} <\frac{\tau_{\rm a,out}}{\tau_{\rm 0,out}} <1.23 \frac{\tau_{\rm a,in}}{\tau_{\rm 0,in}}.
\end{align}
While the scaled timescales follow the thresholds in approach~(i), there is no specific correlation between the values of $\xi$s and the final configuration.
%==================================== nbody-model 3, approach 3===================================================
\begin{figure} 
	\includegraphics[width=\columnwidth]{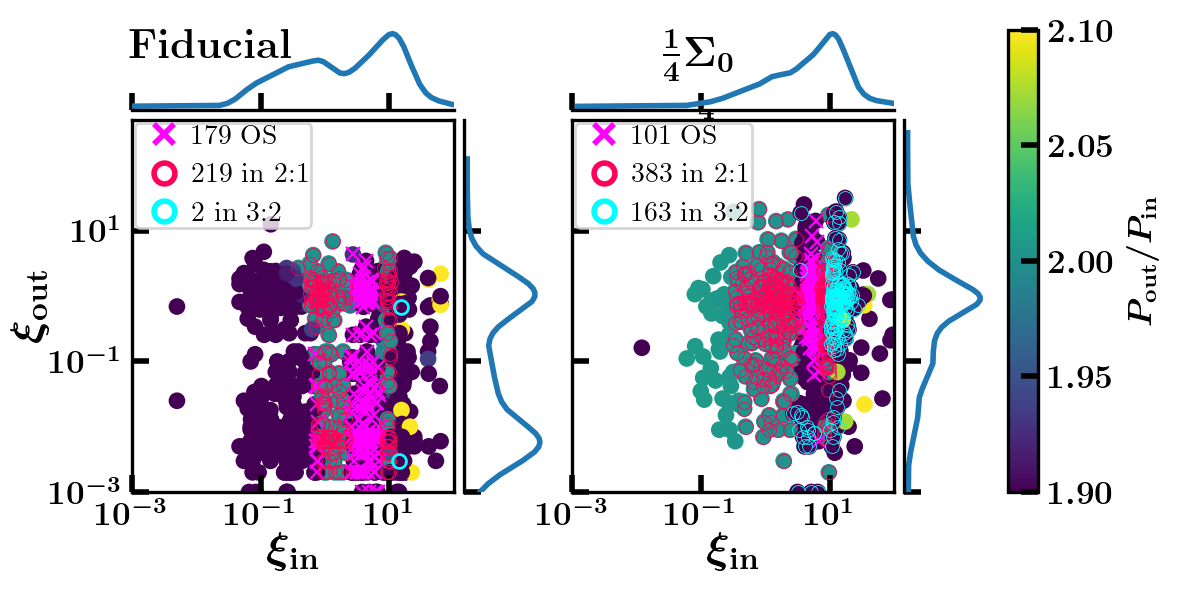}
	\caption{Orbital period ratio of the outer to the inner planet for approach~(iii) with scaled $\tau_{\rm e}$. The horizontal/vertical axes are $\xi$ parameters for the inner/outer planet. All of the overstable~(OS) models are marked with magenta circles. The simulations with final state in a resonance configuration are highlighted with the same colors as in Fig.~\ref{fig:nbody_model1}. The side panels show the distributions used for drawing $\xi$s that are driven from the hydrodynamical simulations. The vertical axes in these panels are not important as they all are scaled to 1.}
	\label{fig:nbody_model3}
\end{figure}

\begin{figure} 
	\includegraphics[width=\columnwidth]{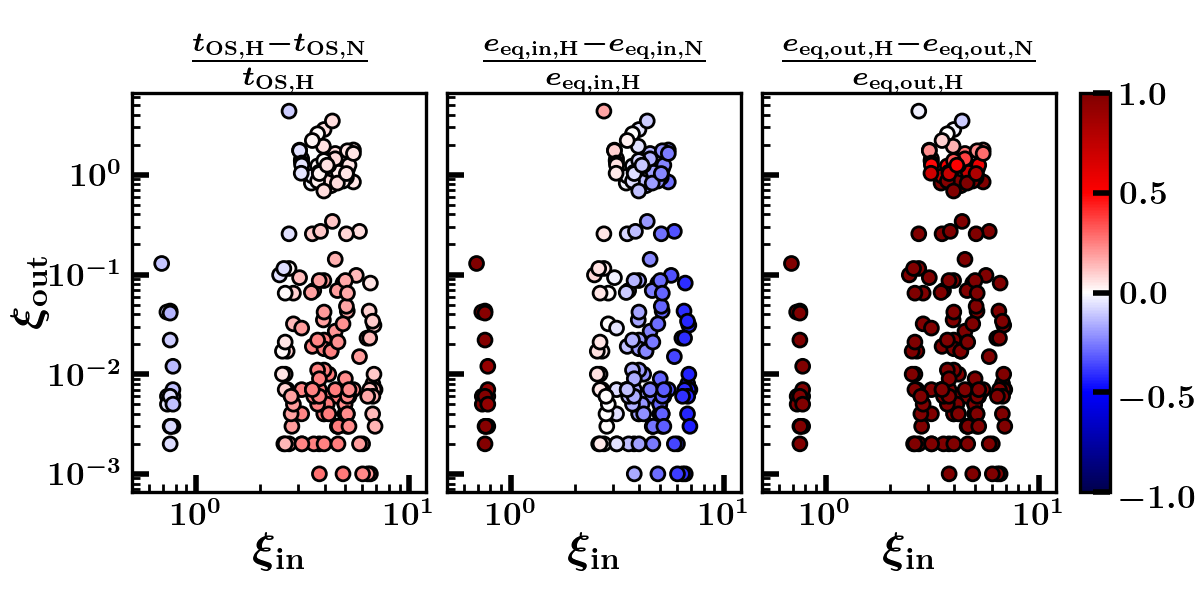}
	\caption{Relative difference between the outcome of fiducial hydrodynamical and the overstable systems in our third N-body approach. Panels from left to right show the overstability duration, inner and outer planet's equilibrium eccentricities. Blue color corresponds to smaller value in the hydrodynamical simulation and red demonstrates the opposite.}
	\label{fig:nbody_model3_oss}
\end{figure}

There are several overstable systems in the approach~(iii): 179 systems ($\sim 18\%$) in the fiducial and 101 systems ($\sim 10\%$)  in the $\nicefrac{1}{4}\Sigma_0$ model.  221 systems ($\sim 22\%$) of the fiducial simulations are  captured and stayed in a MMR. This number is about 2.5 times bigger (546 systems $\sim 50\%$) for the $\nicefrac{1}{4}\Sigma_0$ model. Fig.~\ref{fig:nbody_model3} indicates that in both models, $\xi_{\rm in}$ is the key parameter in determining the fate of the system. The 2:1 resonance happens if $ 0.5 \lesssim \xi_{\rm in} \lesssim 12$,  but the systems with $2.8 \lesssim \xi_{\rm in} \lesssim 6.5$ and few models with $\xi_{\rm in} \sim 0.7$ become overstable in the fiducial model. Similarly, in the  $\nicefrac{1}{4}\Sigma_0$ model, systems with $0.3 \lesssim \xi_{\rm in} \lesssim 10$ are captured in 2:1 resonance but those with $3.4 \lesssim \xi_{\rm in} \lesssim 7$ undergo overstability. Except few cases, almost all of the simulations that stay in 3:2 resonance have $ \xi_{\rm in} \gtrsim 10$. These results imply that overstability needs a moderated value of eccentricity damping for the inner planet.
In Fig.~\ref{fig:nbody_model3_oss}, we compare the overstability duration, inner planet and outer planet's equilibrium eccentricities between our fiducial hydrodynamical simulation and the third approach overstable systems. For obtaining the equilibrium eccentricities, $e_{\rm eq}$, we simply averaged the eccentricity of the planet during the overstable time. The overstability durations in the N-body simulations with $\xi_{\rm out} \gtrsim 0.5$ deviate less than 10\% from the hydrodynamical simulation. For the smaller values of $\xi_{\rm out}$, if $\xi_{\rm in}$ is larger than $\sim 2$, the overstable time becomes shorter than the hydrodynamical simulation and the opposite holds for the smaller values of $\xi_{\rm in}$. The difference is not more than 40\%. The middle panel shows that $\xi_{\rm in} \sim 2 - 3$ produces similar inner planet's equilibrium eccentricity with the hydrodynamical simulation. Likewise to the inner planet, $\xi_{\rm out} \sim 2 - 3$ results in similar outer planet's eccentricity to the hydrodynamical model. The values slightly under or over $\sim 3$ could greatly decrease or increase the outer planet's eccentricity. Therefore, while in the overstable systems, varying $\xi$s does not change the overstability duration, it can greatly affect the equilibrium eccentricities.

%====================================nbody model2: approach 4=================================================== 
\begin{figure} 
	\includegraphics[width=\columnwidth]{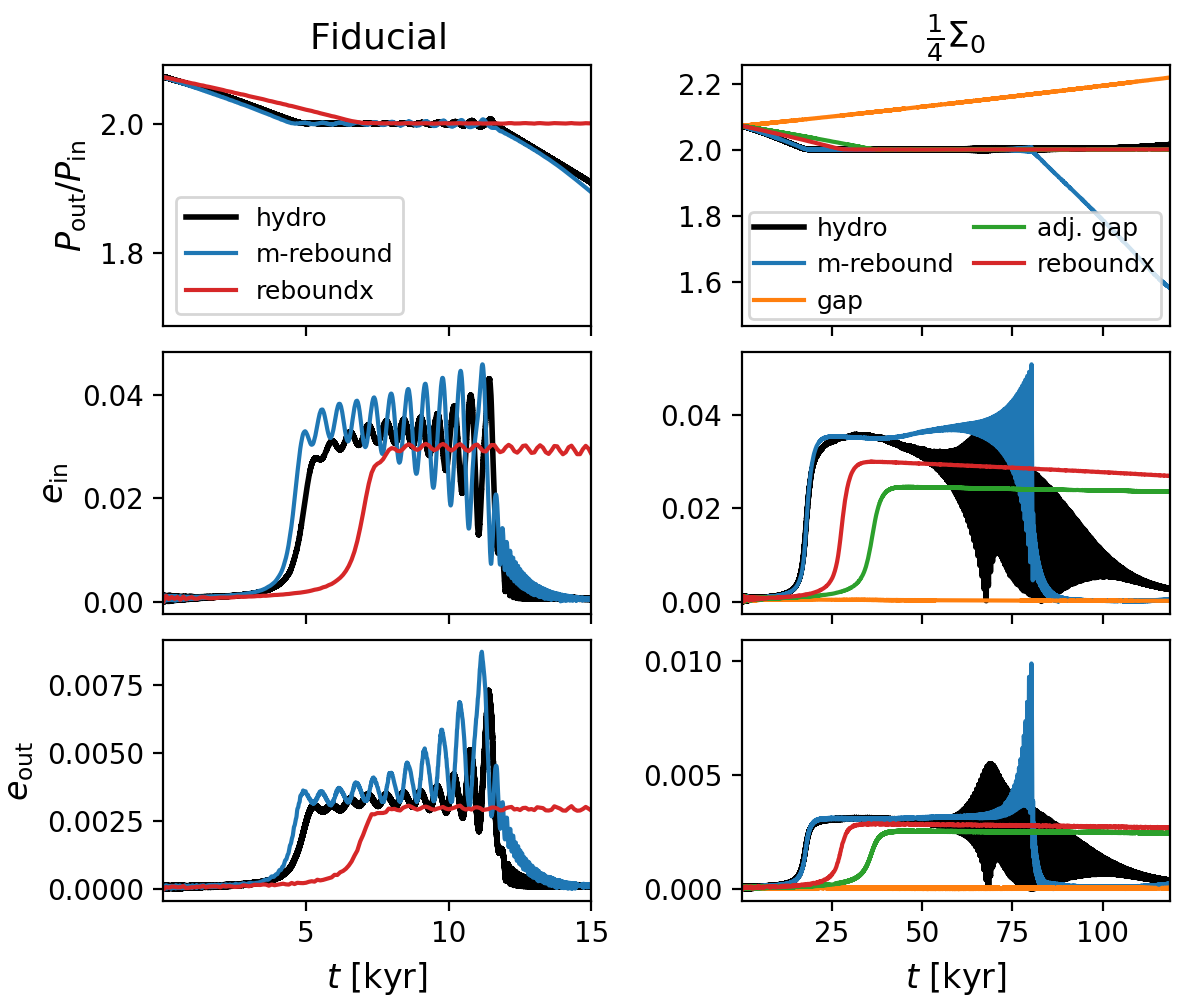}
	\caption{Panels from top to bottom respectively show the orbital period ratio of the outer to inner planet, the inner planet's and the outer planet's eccentricities for the fiducial model~(left) and $\nicefrac{1}{4}\Sigma_0$ model~(right). \texttt{m-rebound}, \texttt{gap}, and \texttt{adj.~gap} respectively represent modified version of the \texttt{REBOUND} code that is used in this work, gap model included exactly as in \citet{2018Kanagawa}, and adjusted gap model as explained in the text.}
	\label{fig:nbody_model2}
\end{figure}

In the forth approach, in which the timescales are calculated every timestep using the torque on the planets, we found a very good agreement between the result of the N-body and hydrodynamical simulations for the fiducial model if the \texttt{REBOUND} code with the migration method in \cite{2021A&A...AK} is used~(Fig.~\ref{fig:nbody_model2}). However, the same method for $\nicefrac{1}{4}\Sigma_0$ model gives different results. The difference can be related to the partial gap-opening in the $\nicefrac{1}{4}\Sigma_0$ model. The method in \cite{2021A&A...AK} uses the type-I migration without any gap-opening. As indicated in \cite{2024A&A...Zahra}, the planets migration in the fiducial model is type-I while the $\nicefrac{1}{4}\Sigma_0$ is between type-I and type-II due to the partial gap opening. For this later model, we performed two more simulations taking into account the effect of gap opening in the torques as suggested in \cite{2018Kanagawa}. In one of these simulations, we decrease the coefficient of $K$ in the relation~(10) of \cite{2018Kanagawa} from \num{0.04} to \num{0.0001} to mimic a shallow partial gap using less surface density drop at the planet's orbit. The results can be compared in the right column of Fig.~\ref{fig:nbody_model2}. Type-I migration results in overstability after about \num{75000} \si{\yrs}, considering full gap opening results in diverging migration, and reducing the gap depth leads to a later overstability~(after \num{110000} \si{\yrs}). These results suggest that type-I migration is modified by the partial gap opening in the $\nicefrac{1}{4}\Sigma_0$ model.

The type-I migration approach in \texttt{REBOUNDx} results in 2:1 resonance for both models. It implies that the planets migration are slower in these cases. The reason is that the type-I migration in \texttt{REBOUNDx} is modelled using \cite{2008A&A...CN} where the torque relation is taken from \cite{2002ApJ...565.1257T}, which calculates the torque for an isothermal disc without including the saturation of co-rotation torque. This causes the positive co-rotation torque, as explained in \cite{2024A&A...Zahra}, slows down the migration of the planets, that results in capturing and staying in the 2:1 resonance.

%========================================Section: Discussion===========================================
\section{Discussion} \label{sec: discussion}
%\subsection{Dynamical Drivers of Alternating Planetary Migration}
\subsection{Origin of the sign reversals in migration and eccentricity damping timescales} \label{subsec: Bifurcation timescale}

Our analysis shows that the migration and eccentricity damping timescales are not single--valued quantities but instead exhibit oscillatory behaviour with alternating positive and negative values (Fig.~\ref{fig:tau_ae_column}). Nevertheless, the long-term evolution still corresponds to net inward migration.

To investigate the origin of these sign reversals, we performed a high-temporal-resolution simulation over a short interval during resonance. The corresponding evolution of the fiducial model is presented in Appendix~\ref{appendix: extra simulations} (Fig.~\ref{fig:snap_highreso_during_in_out}). Although the general evolution of the semi--major axes indicates inward migration, short-timescale oscillations are superimposed on this trend. These oscillations produce increases and decreases in the semi--major axes, corresponding to local sign reversals in the migration timescale.

Additional test simulations (see fig.~\ref{fig:torq_power_ecc}) indicate that, in the absence of mutual gravitational interactions and eccentricity excitation, the migration and damping timescales remain single-signed. In contrast, systems with non--zero eccentricity and planet--planet interactions exhibit alternating positive and negative timescales. This suggests that the observed variability is closely linked to eccentricity and the resulting variations in the disc torque. Previous studies have shown that eccentric planets can modify the disc torque and migration behaviour through changes in the disc response and spiral structure \citep[e.g.][]{2000MNRAS...PL, 2007A&A...CDKN, 2013ApJ...778....7B}.

The azimuthally averaged surface density profiles shown in Appendix~\ref{appendix: extra simulations} (Fig.~\ref{fig:snap_highreso_during_in_out}) further suggest that close planetary conjunctions and spiral-arm interactions may contribute to the torque variability. In particular, the spiral wake generated by one planet can pass through the co-orbital region of the other planet, potentially producing additional perturbations in the torque. Similar modifications of the disc structure and torque behaviour for eccentric planets have previously been reported by \cite{2007A&A...CDKN}.

These results suggest that migration and eccentricity damping in resonant multi-planet systems are intrinsically time-variable processes rather than strictly constant quantities.

\subsection{Viscosity dependence and implications for resonant overstability}
Our results show that the dynamical evolution of resonant planetary systems depends on the disc viscosity. In our models, overstability occurs for $\alpha_{\nu}=10^{-6}, 10^{-5}$, and $10^{-4}$, whereas the model with $\alpha_{\nu}=10^{-3}$ exhibits limit-cycle behaviour, with the planets remaining in resonance \citep{2024A&A...Zahra}.

In protoplanetary discs, angular momentum transport is governed by complex physical processes whose relative importance may vary with radius \citep{Bai_2026}. Viscous disc models have also been used to infer radial variations in the viscosity profile from observations \citep{Isella_2009}, while model-based estimates of the effective $\alpha$-viscosity span a broad range across different discs \citep{Rafikov_2017}. These studies provide context for examining how the viscosity dependence found in our simulations relates to conditions in protoplanetary discs.

If the effective viscosity varies with radius, the disc conditions governing resonant evolution may also vary with orbital location. For example, viscosity in a disc with a constant $\alpha_\nu$ and temperature gradient $\beta=-\frac{d \log T}{d \log r}  < 1.5$ increases with radius. Therefore, overstability is more expected to happen in the inner part of the disc. On the other hand, overstability is unlikely to happen the inner edge of the discs where MRI is active and $\alpha_\nu \sim 10^{-2}$ \citep{2011ApJ...736..144B}. In addition, other mechanisms might intervene and change the outcome of overstability~\citep{2022ApJ...925...38N,batygin2026suppression}. However, the implications of such radial variations for resonant overstability remain to be investigated using simulations that account for the radial structure of the disc.

\subsection{Suggested scheme(s) for N-body simulations} \label{sec:which-nbody}
In Sec.~\ref{sec:nbody}, we examined various approaches for choosing the migration and e-damping timescales, from constant values to more realistic time-varying timescales based on the torque from the disc. Because there was absolutely no agreement between the results of constant and scaled timescale(s) simulations with their hydrodynamical counterpart in the fiducial model, we do not recommand these approaches for modelling resonance overstability with N-body simulation. In the first two approaches, $\xi$ and the e-damping timescale play no notable role in the final configuration and the most determining parameter in the fate of the system is the ratio of migration timescales. We should note that in these two approaches we assumed the random values from the used distributions are equal to the orbital averaged quantities in the N-body simulations.

In the third approach, $\xi$ of the inner planet is the most important parameter and $3 \lesssim \xi \lesssim 7$ results in overstability. Our results show that $\xi$ does not necessarily equal to unity as mentioned in \cite{2014AJ...GS} and can vary over a large span. The value of $\xi=10$ suggested by \cite{batygin2026suppression} for suppression of overstability is well above our derived span. However, \cite{2021A&A...AK} found $\xi$ is between 5--15 for overtable system. The difference between the results of our study and \cite{2021A&A...AK} can be attributed to the surface density profiles that deviate from power-law relation at the disc inner edge. In addition, $\xi_{\rm out} \gtrsim 0.5$ produces similar overstability duration as in the hydrodynamical model, $\xi$ for the inner and outer planet should be only between 2--3 to have similar equilibrium eccentricities in the N-body simulations to the hydrodynamical ones. This values are smaller than the expected $\xi$ obtained by the relations in \cite{2008A&A...CN} for our disc parameters. 

In the torque calculated timescale simulations, we acquired the best agreement for the fiducial model when the method in \cite{2021A&A...AK} is used. In other words, changing the co-rotation torque by the saturation \citep{2011MNRAS.410..293P} and planet's eccentricity \citep{2014MNRAS.437...96F} play a crucial role in finding agreement between N-Body and hydrodynamical results. However, in the $\nicefrac{1}{4}\Sigma_0$ model, we had to include a partial gap-opening effect to reduce the type-I migration rates and found a closer result to the hydrodynamical model.

In the N-body simulations presented in Sec.~\ref{sec:nbody}, we draw the random quantities from distributions based on the outputs of the hydrodynamical models. Therefore, the selection are almost independent. However, the timescales in the hydrodynamical simulations are related that means if we choose a random migration timescale for the inner planet, the rest of timescales are determined by the dynamics and configuration of the system at that specific moment. To examine how much this dependency affect the final results, we repeated our first three approaches but we used the corresponding parameters after drawing the first quantity. The overall results do not change~(see Appendix~\ref{appendix: dependent} for more details).
%=====================================================================================================
\section{Summary and conclusions} \label{sec: conclusion}

In this work, we investigated migration and eccentricity damping timescales in overstable resonant planetary systems using hydrodynamic simulations and assessed the validity of several commonly adopted N-body prescriptions. Building on the hydrodynamic models presented in \cite{2024A&A...Zahra}, we focused on the evolution of the migration and damping timescales themselves. We examined their role in reproducing resonant and overstable behaviour.

Our main findings can be summarised as follows:

\begin{enumerate}
	
	\item Both migration and eccentricity damping timescales are intrinsically time-dependent quantities. Rather than remaining constant throughout the evolution, they exhibit substantial temporal variability and sign reversals associated with resonant evolution, eccentricity oscillations, and variations in the disc torque.
	
	\item After factoring out the explicit dependencies on planetary mass, disc surface density, and aspect ratio represented by the adopted Type-I reference scaling, the resulting scaled migration timescales obtained from different models do not appear to collapse onto an identical curve within the explored parameter range.
	This indicates that additional physical effects associated with the evolving planet-disc interaction remain important in determining the migration response, including changes in disc structure, gap formation, spiral-wave structures, and gas depletion.
	
	\item The commonly adopted relation between migration and eccentricity damping timescales is not recovered by our hydrodynamic measurements. The parameter $\xi$ differs from unity and varies between the inner and outer planets, indicating that the simple analytical scaling frequently adopted in N-body studies does not accurately describe the behaviour of our overstable systems.
	
    \item We found that a careful treatment of torque calculation in an N-body code can reproduce the hydrodynamic evolution of resonant planetary systems undergoing type-I migration. In particular, the torque prescription should account for the dependence of the co-rotation torque on both co-orbital saturation and planetary eccentricity. However, when the disc and planetary properties allow for partial gap opening, the assumptions underlying pure type-I migration are no longer valid. We suggest either our third approach with using proper values for $\xi$s, or our forth approach with corrected recipes  for (a)~ the saturation and eccentricity effect on the torque, and (b)~type-I migration with partial gap specially if the viscosity of the disc is low or the migration rate of the planet is very slow such that the planet has enough time to open a partial gap.

\end{enumerate}

Taken together, these results suggest that migration and eccentricity damping in resonant multi-planet systems should be regarded as evolving dynamical quantities rather than fixed parameters. While constant timescale prescriptions remain useful for simplified studies, they may not adequately capture the complex disc-driven evolution responsible for resonance maintenance, resonance escape, and overstability. Our results therefore support the use of torque-based migration models when attempting to reproduce hydrodynamic behaviour within computationally efficient N-body calculations.

An important next step can be to determine whether the variability identified here modifies the stability boundaries predicted by existing analytical theories of resonant overstability

\section*{Acknowledgements}

SA and ZA gratefully acknowledge the research support by Iran National Science Foundation (INSF) under project No 4049507.
A.B.T.P. and A.Z. acknowledge funding from the European Union under the European Union's Horizon Europe Research and Innovation Programme 101124282 (EARLYBIRD). Views and opinions expressed are, however, those of the authors only and do not necessarily reflect those of the European Union or the European Research Council. Neither the European Union nor the granting authority can be held responsible for them.
The N-body part of the calculation was conducted at the Sci-HPC center of the Ferdowsi University of Mashhad. 
The authors acknowledge support by the High Performance and Cloud Computing Group at the Zentrum für Datenverarbeitung of the University of Tübingen, the state of Baden-Württemberg through bwHPC and the German Research Foundation (DFG) for funding under "Project number 455787709" (bwForCluster BinAC 2).
Z.A. thanks Thomas Rometsch for his helpful discussions and Yousef Sadeghi for conducting some of the initial tests for the N-body simulations. We are grateful to Ruobing Dong who carefully refereed this manuscript and helped to improve it considerably.

%%%%%%%%%%%%%%%%%%%%%%%%%%%%%%%%%%%%%%%%%%%%%%%%%%
\section*{Data Availability}

The data of the simulations in this manuscript will be shared on reasonable request to the corresponding author.



\bibliographystyle{mnras}
\bibliography{example} % if your bibtex file is called example.bib

%%%%%%%%%%%%%%%%%%%%%%%%%%%%%%%%%%%%%%%%%%%%%%%%%%

%%%%%%%%%%%%%%%%% APPENDICES %%%%%%%%%%%%%%%%%%%%%

\appendix 

\section{Azimuthally averaged surface density for all models} \label{appendix: azimutally averaged surface density}

\begin{figure*} 
	\centering
	
	\includegraphics[width=\textwidth]{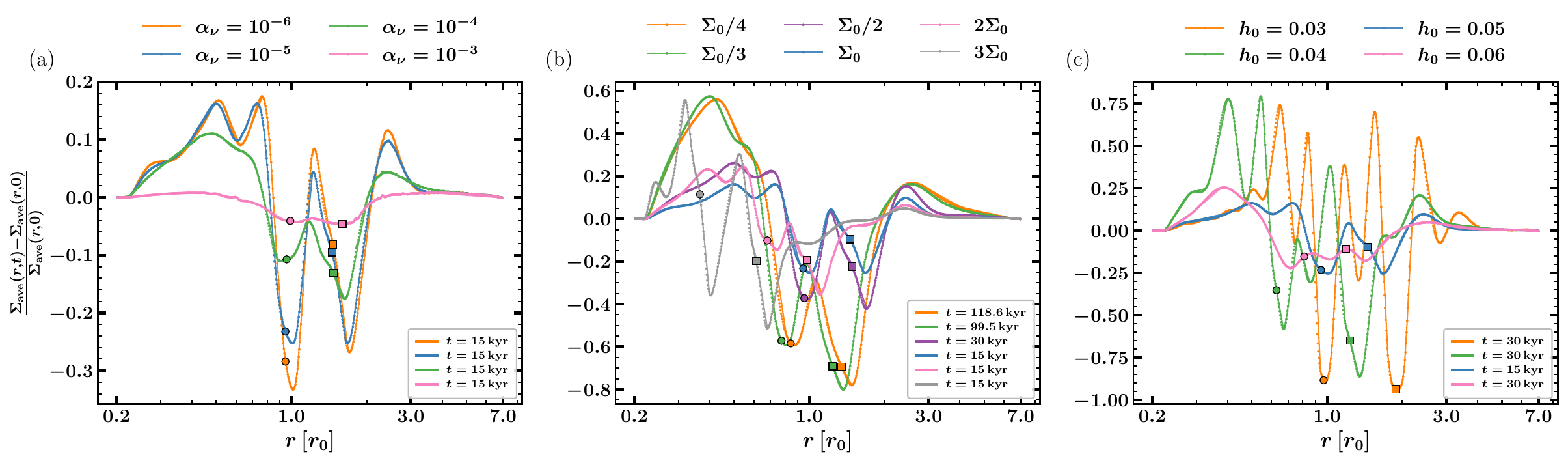}
	\caption{Relative change in the azimuthally averaged surface density profile for
		models with different (a) viscosity, (b) initial surface density, and
		(c) aspect ratio.
		Circles and squares mark the radial positions of the inner and outer planets, respectively. The planet positions are evaluated at the same output time as the corresponding surface density profile. As a reminder, in each model sequence, the other disc parameters are kept at their fiducial values.}
	\label{fig:azi_combined}
\end{figure*}

In Fig.~\ref{appendix: azimutally averaged surface density}, we present the relative change in the azimuthally averaged surface density profiles for all models at the final simulation time available for each run. Here, $\Sigma_{\rm ave}(r,t)$ denotes the surface density averaged over the azimuthal direction at radius $r$ and time $t$, while $\Sigma_{\rm ave}(r,0)$ denotes the corresponding azimuthally averaged surface density in the initial state.
	
For the viscous models and the surface density models (excluding $\Sigma_{0}/4$, $\Sigma_{0}/3$, and $\Sigma_{0}/2$), this corresponds to $t = 15$~kyr. The $\Sigma_{0}/4$ model reaches $t = 118.6$~kyr, the $\Sigma_{0}/3$ model reaches $t = 99.5$~kyr, and the $\Sigma_{0}/2$ model reaches $t = 30$~kyr. For the aspect ratio models, most are shown at $t = 30$~kyr, except for the $h_0 = 0.05$ model, which is shown at $t = 15$~kyr.
	
The curves are shown together for compactness and to avoid unnecessary duplication of similar figures. Therefore, they are not intended for a direct comparison between different models at the same simulation time. Instead, for each model, the corresponding relative surface density profile and gap structure are compared with the scaled migration timescales of the same model presented in Figs.~\ref{fig: alphas}-\ref{fig: hs}. Since the simulations have different duration, we use the latest available time in each run to examine the latest disc state accessible in that simulation. This provides the disc structure at the latest available stage of each simulation for interpreting the evolution of the scaled migration timescales of the corresponding model.

\section{Distribution of the $\xi$ parameter during resonance} \label{app:violet_plot}

%=========================================Figure: xi_violin_plot=======================================
\begin{figure*} 
	\centering
	\includegraphics[width=\textwidth]{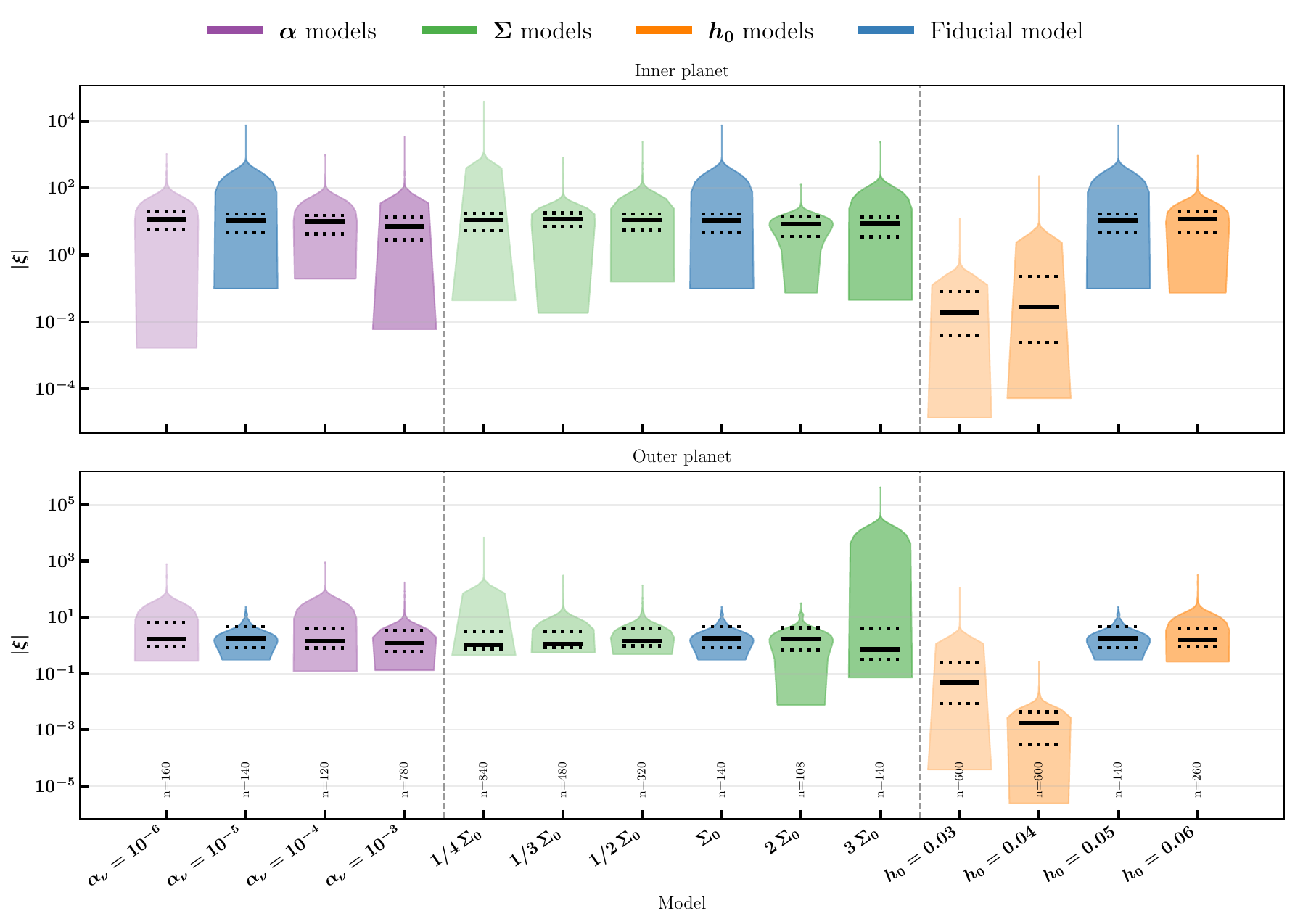}
	\caption{Distribution of $|\xi|$ during the resonant phase for the inner (top) and outer (bottom) planets in all disc models. For each simulation, the median and the $16$th--$84$th percentiles are computed over the resonant interval identified for that model, yielding different sample sizes ($n$ denotes the sample size for each model). The fiducial model (blue) is repeated in each group to facilitate comparison with the viscosity (purple), surface density (green), and aspect ratio (orange) models. The horizontal solid and dashed lines indicate the median and the $16$th--$84$th percentiles, respectively.
	}

	\label{fig:xi_violet_plot}
\end{figure*}

Figure~\ref{fig:xi_violet_plot} presents the distributions of $|\xi|$ for all simulations using violin plots. While Figs.~\ref{fig:xi_alphas}--\ref{fig:xi_hs} emphasize the temporal evolution of $\xi$, the violin plots summarize the statistical behaviour of each model over its resonant interval. In particular, they illustrate the characteristic values of $|\xi|$, the width of the distributions, and the occurrence of extreme excursions, thereby facilitating a direct comparison among different disc models.
%====================================================================================================== 

%========================= N-body integrators comparison ===========================
\section{Comparing \texttt{IAS15} and \texttt{WHFAST} integrators in the \texttt{REBOUND} code} \label{appendix: integrators}
\begin{figure*} 
	\centering
	\includegraphics[width=\textwidth]{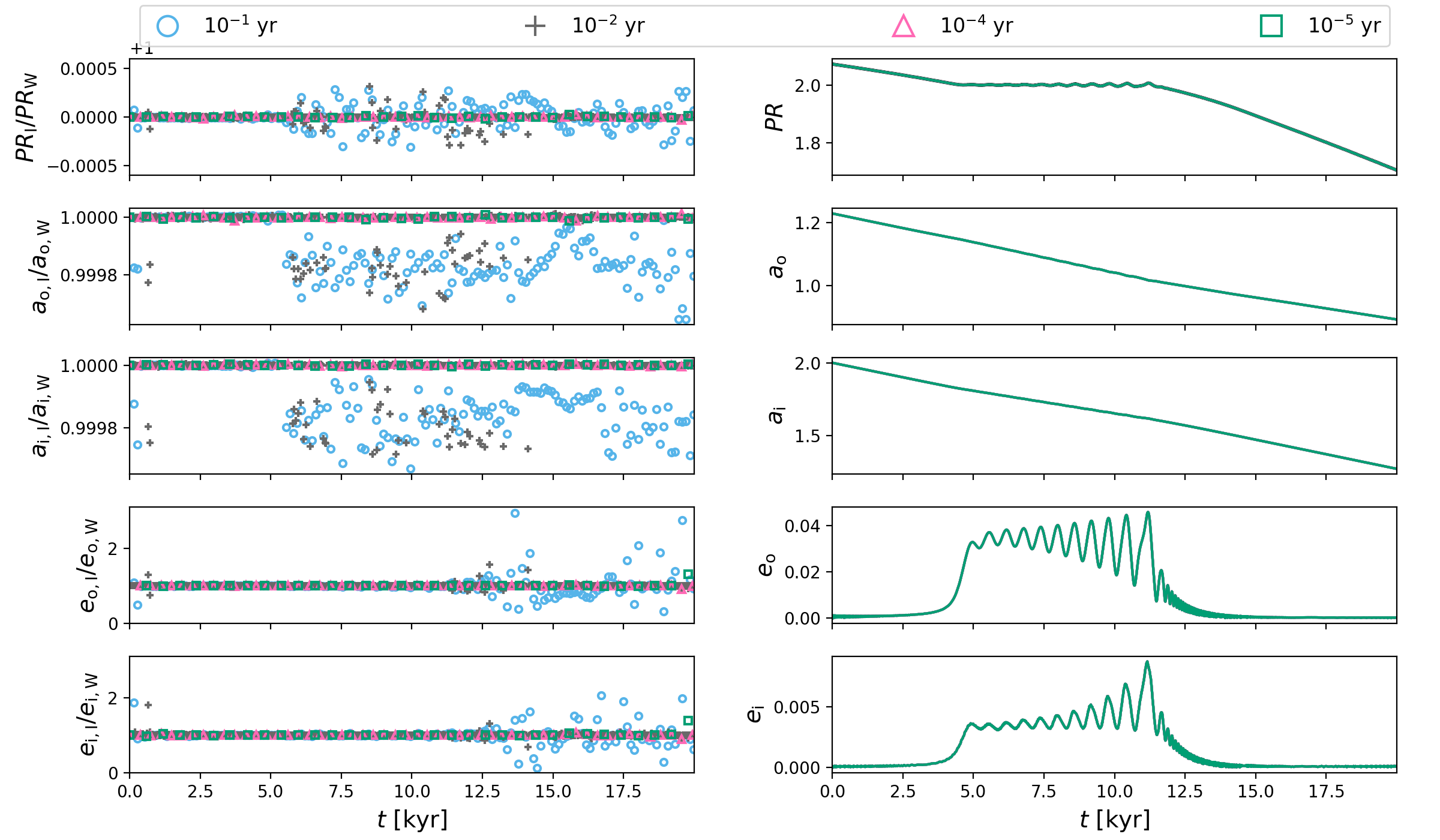}
	\caption{Comparing the outputs of \texttt{REBOUND} simulations in which \texttt{WHFAST} integrator with different time~steps are used with the \texttt{IAS15} integrator. In these plots, PR stands for period ratio of the outer to the inner planet, and the subscript I/W represents \texttt{IAS15}/\texttt{WHFAST}. The legend show the time~steps used in the \texttt{WHFAST} simulations.}
	\label{fig:integrators}
\end{figure*}

In Fig.~\ref{fig:integrators}, we present the outputs of the \texttt{REBOUND} simulations for the fiducial model in which the \texttt{WHFAST} integrator with constant time~steps are applied against an identical simulation that uses \texttt{IAS15} adaptive time~step integrator. In the right panels, the \texttt{IAS15} results are shown in black and the \texttt{WHFAST} simulations in colours. No difference is recognisable. In the left panels, we plot the ratio of \texttt{IAS15} to \texttt{WHFAST} outputs. For the \texttt{WHFAST} simulations which have time~step $dt = 10^{-4}$ and $10^{-5}$ of a year, the results are almost identical to the one with \texttt{IAS15}. We note that this conclusion should not be generalised to the simulations that close encounters are interested, as they could need time~steps smaller than $dt = 10^{-5}$~yr.

%=====================================================================================================

\section{Additional Analysis of the Sign Reversals in Migration and Eccentricity Damping Timescales} \label{appendix: extra simulations}
\begin{figure*} 
	\centering
	\includegraphics[width=\textwidth]{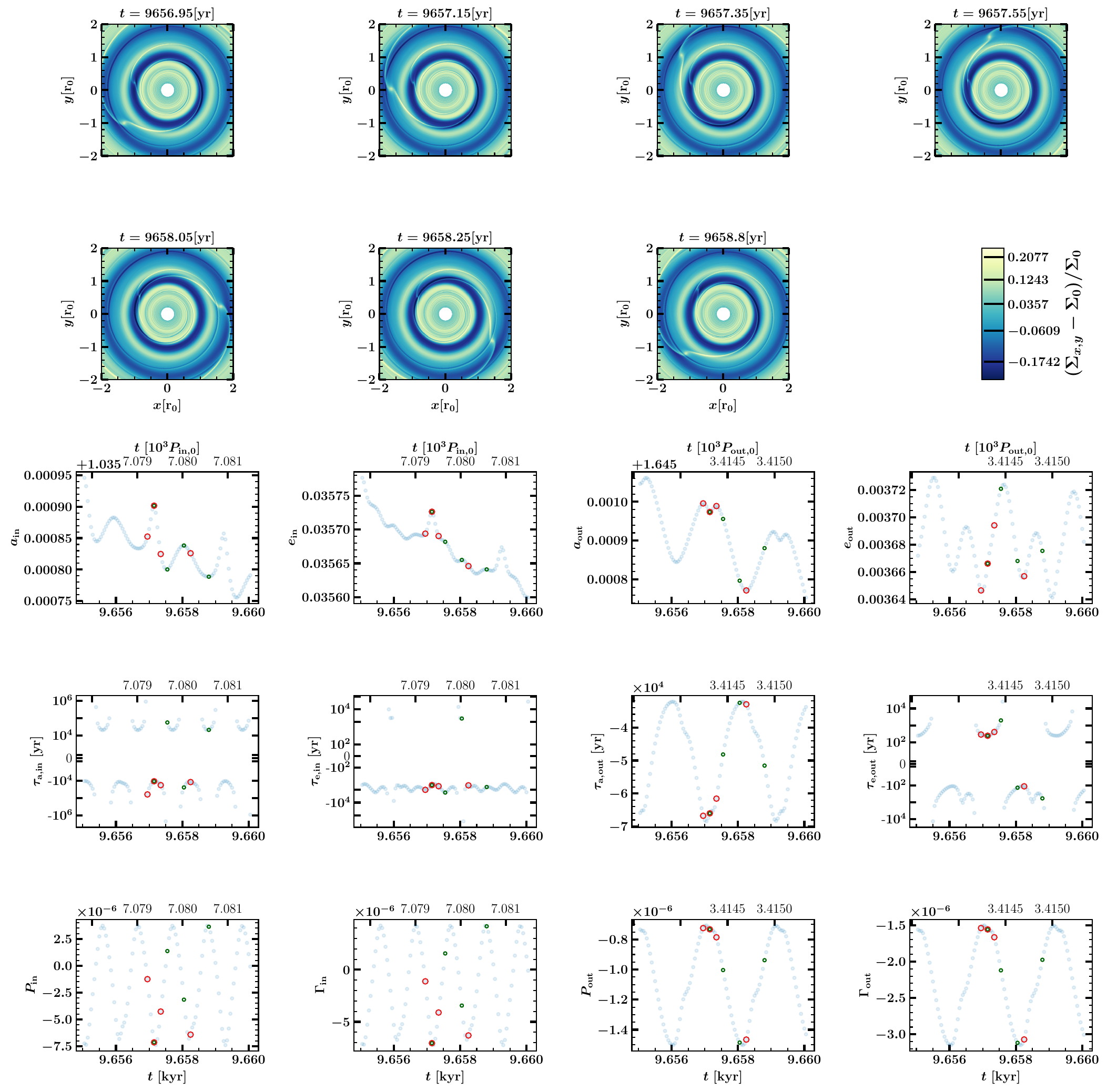}
	\caption{Snapshots of the normalised surface density perturbation are shown at the times marked by the colored circles in the lower panels. These times correspond to successive extrema of the semi-major axis during one representative oscillation cycle. Small green circles mark the selected extrema of the inner planet, while large red circles mark those of the outer planet. The snapshots in the first and second rows correspond to these marked epochs, in chronological order.
 In rows three to five, The first two columns (from the left) refer to the inner planet, and the next two to the outer planet. The second row shows the semi-major axes and eccentricities; the third row shows the migration and eccentricity damping timescales; and the last (fifth) row shows the power and torque exerted on the planets by the disc. All plots correspond to epochs during planetary resonance in the fiducial model.}
	\label{fig:snap_highreso_during_in_out}
\end{figure*}

Figure ~\ref{fig:snap_highreso_during_in_out} shows a high-temporal-resolution segment of the fiducial simulation during resonance, highlighting the short-timescale variability associated with the sign reversals of the migration and eccentricity damping timescales discussed in Section ~\ref{subsec: Bifurcation timescale}.

%================================== figure: torques and powers========================================

To demonstrate that the sign reversal of the migration and eccentricity damping timescales is caused by both planet--planet interactions and eccentricity, we performed a set of additional simulations.

As reference cases, we carried out simulations with a single migrating planet. One simulation contained a $5,M_{\earth}$ planet located initially at 1.23 AU (hereafter the |Inner Single Planet model), while another contained a $10,M_{\earth}$ planet located initially at 2.0 AU (hereafter the Outer Single Planet model). For both models, we considered initial eccentricities of $e=0$ and $e=0.1$.

For the single-planet simulations with circular orbits ($e=0$), no sign reversal was observed in either the torque or the power. This indicates that, in the absence of both planet--planet interactions and orbital eccentricity, the corresponding timescales do not exhibit sign reversal. In contrast, sign reversals are present in the torque and power for single planets on eccentric orbits ($e=0.1$), demonstrating that eccentricity alone can produce this behaviour.

To isolate the effect of gravitational interactions, we also performed a two-planet simulation in which planet--planet interactions were switched off. Since the planets remained on circular orbits, no sign reversal was observed. However, in the $h=0.03$ model, where the planets interact gravitationally but their eccentricities remain negligible, sign reversals are present. Finally, in the fiducial model, where both planet--planet interactions and eccentricity are present simultaneously, sign reversals are also observed.

\begin{figure*} 
	\centering
	\includegraphics[width=\textwidth]{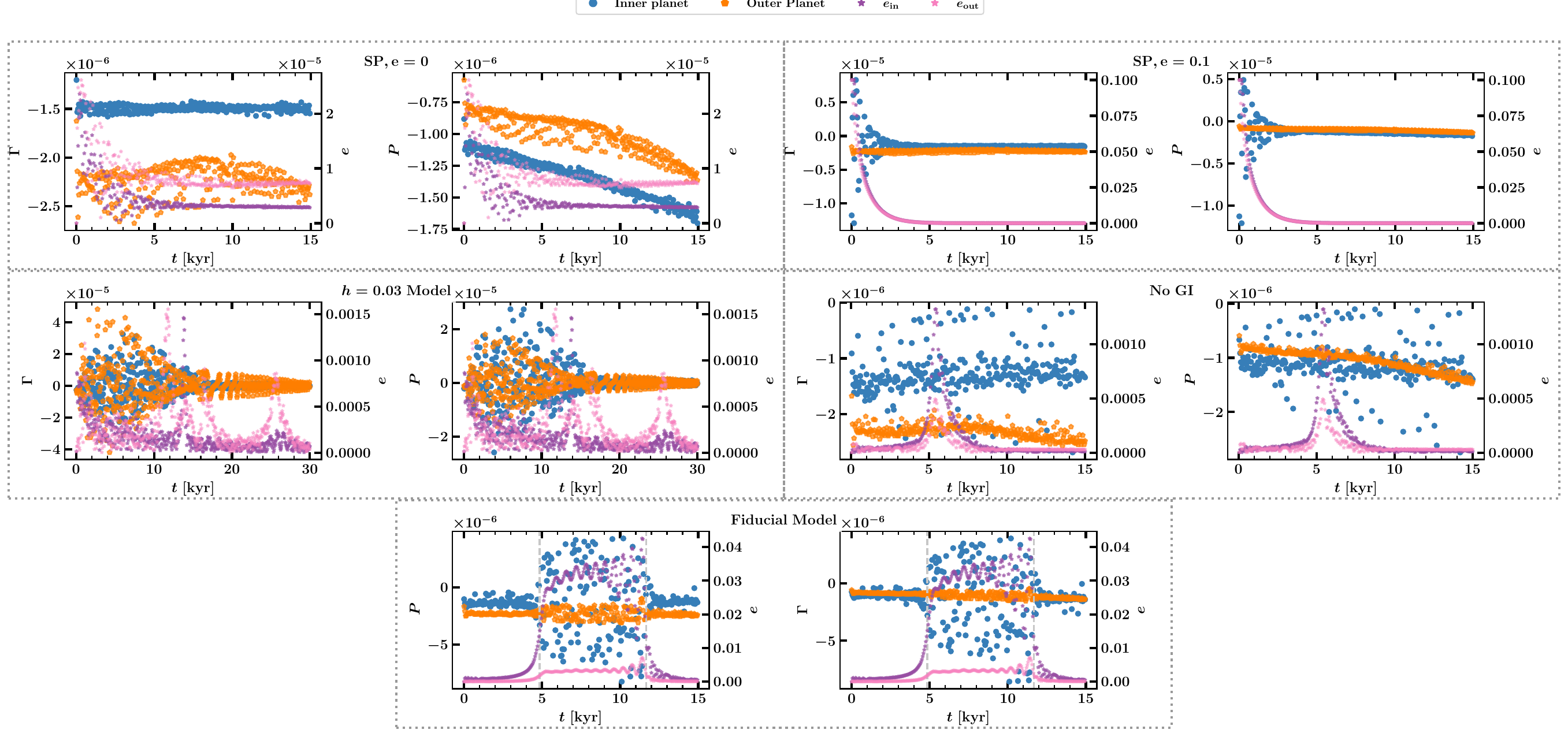}
	\caption{Temporal evolution of torque and power for five different models. The first row shows single-planet models (inner planet in blue, outer planet in orange). In the left panel, the initial eccentricity is zero; in the right panel,  $e = 0.1$.
		The second row: the left panel corresponds to the model with $ h_0 = 0.03$, and theright panel shows the fiducial model without mutual planet--planet interactions.
		The third row (last panel) shows the fiducial model (for reference).
		Markers: Purple and pink indicate the eccentricities of the inner and outer planets, respectively. Blue and orange represent the torque/power values for the inner and outer planets, respectively.
	}
	\label{fig:torq_power_ecc}
\end{figure*}

These results are illustrated in Fig.~\ref{fig:torq_power_ecc}.

%================================== figure: torques and powers========================================

%=====================================================================================================
\section{The effect of parameter dependency on the N-body simulations} \label{appendix: dependent}
\begin{figure} 
	\includegraphics[width=\columnwidth]{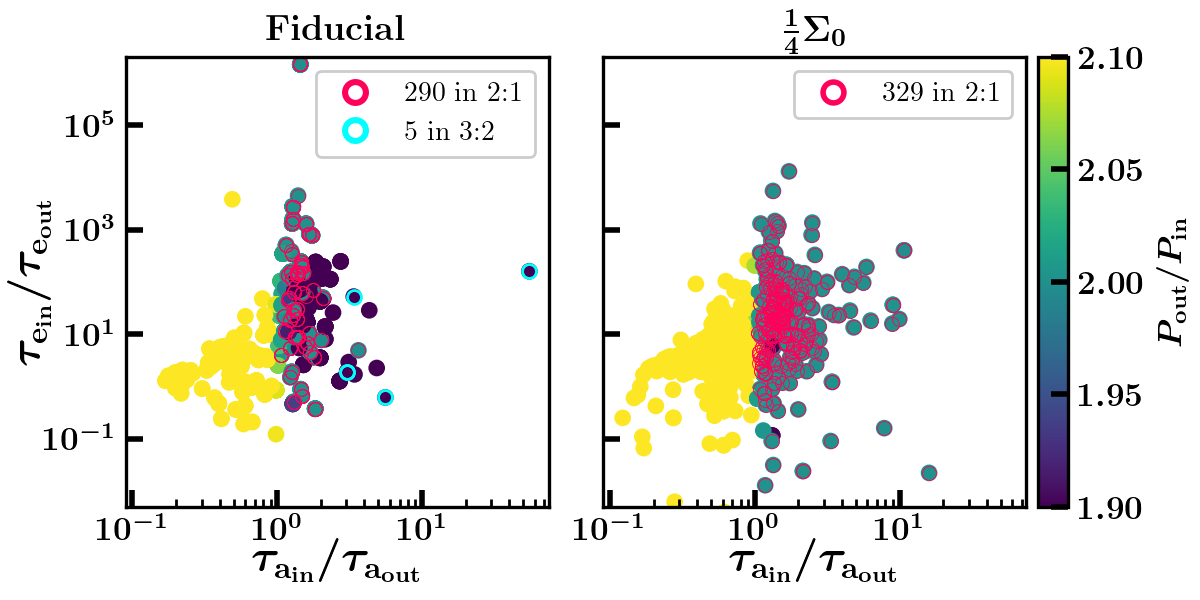}
	\caption{Orbital period ratio of the outer to the inner planet for the simulations in approach~(i) but with taking the dependency of the paramters into account. }
	\label{fig:nbody_model1dep}
\end{figure}

\begin{figure} 
	\includegraphics[width=\columnwidth]{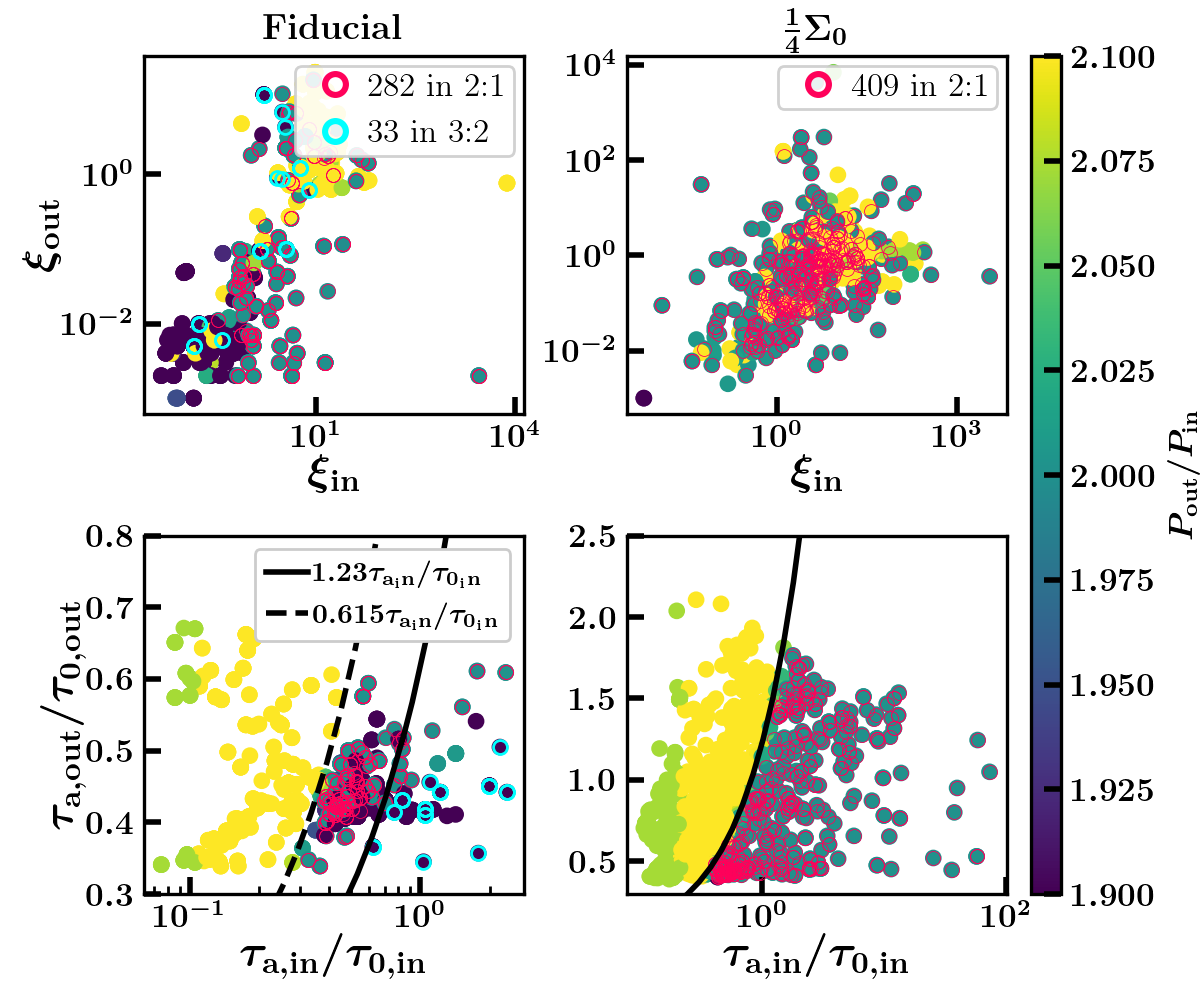}
	\caption{Similar to Fig.~\ref{fig:nbody_model4} but with taking the dependency of the paramters into account.}
	\label{fig:nbody_model4dep}
\end{figure}

\begin{figure} 
	\includegraphics[width=\columnwidth]{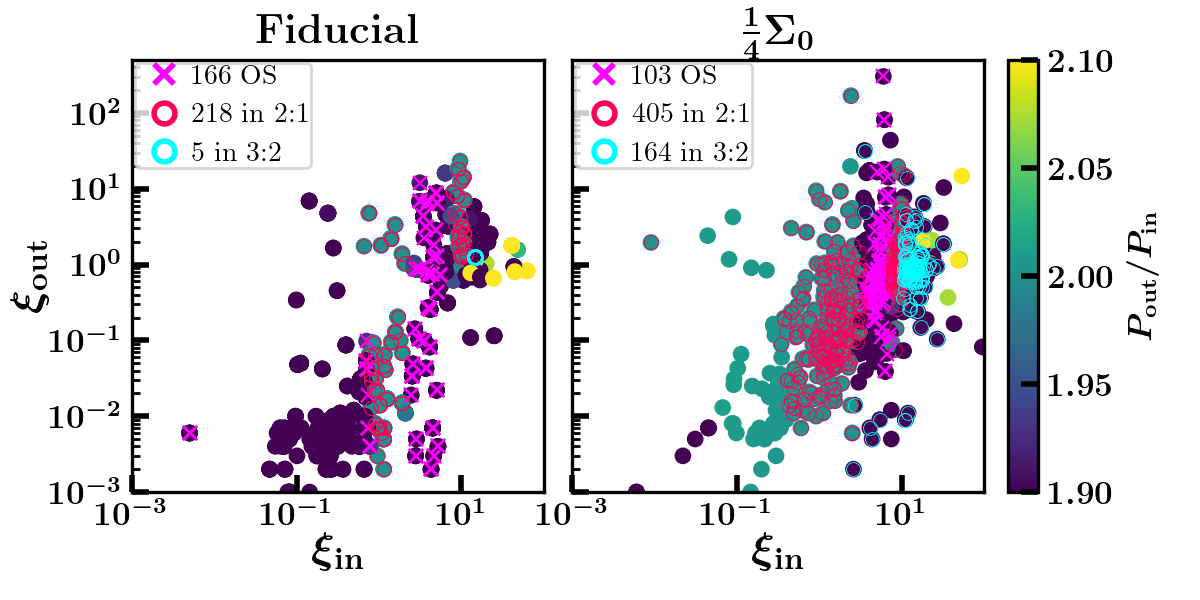}
	\caption{Similar to Fig.~\ref{fig:nbody_model3} but with taking the dependency of the paramters into account.}
	\label{fig:nbody_model3dep}
\end{figure}

We repeated the first three N-body approaches taking the dependency of the parameters into account. After we randomly choose the first parameter from the given distribution as explained in Sec.~\ref{sec:nbody}, the rest of the timescales or $\xi$s are set to its interrelated values. For example, in approach~(i), first we draw a random $\tau_{\rm a,in}$ from its distribution, and then set the other three timescales according to that specific time. Figures.~\ref{fig:nbody_model1dep} to \ref{fig:nbody_model3dep} show that the final results are not much affected by the dependency. However, as in the Sec.~\ref{sec:nbody}, we used the distributions from the results of hydrodynamical simulations, the distributions embed the dependency and the random quantities might not completely be independent.

%%%%%%%%%%%%%%%%%%%%%%%%%%%%%%%%%%%%%%%%%%%%%%%%%%

% Don't change these lines
\bsp	% typesetting comment
\label{lastpage}
\end{document}